\documentclass[%
nofootinbib,
reprint,
superscriptaddress,
amsmath,amssymb,
aps,
prr,
]{revtex4-2}

\usepackage{graphicx}
\usepackage{dcolumn}
\newcommand{\rme}{\mathrm{e}}
\newcommand{\mc}[2]{\begin{tabular}{@{}c@{}}#1 \\ #2\end{tabular}}
\usepackage{nicefrac}
\usepackage{bm}
\usepackage{braket}
\newcommand{\mb}{\mathbf}
\newcommand{\pll}{^{\parallel}}
\newcommand{\pp}{^{\perp}}
\newcommand{\p}{^{\prime}}
\newcommand{\proj}[2]{\left|#1\right>\left<#2\right|}
\newcommand{\uk}{_{\mathbf{k}}}
\newcommand{\kl}{_{\mathbf{k}\lambda}}

\newcommand{\curl}{\boldsymbol{\nabla}\times}
\newcommand{\dive}{\boldsymbol{\nabla}\cdot}
\newcommand{\x}{(\mathbf{x})}
\newcommand{\xr}{\left(\mathbf{x},\mathbf{r}\right)}
\newcommand{\xR}{\left(\mathbf{x},\mathbf{R}_k\right)}
\newcommand{\xpR}{\left(\mathbf{x}^{\prime},\mathbf{R}_k\right)}

\newcommand{\qx}{(\mathbf{x})}
\newcommand{\qxp}{(\mathbf{x}^{\prime})}
\newcommand{\xx}{(\mathbf{x},\mathbf{x}^{\prime})}
\newcommand{\rr}{(\mathbf{r})}
\newcommand{\xp}{(\mathbf{x}^{\prime})}
\newcommand{\xpr}{\left(\mathbf{x}^{\prime},\mathbf{r}\right)}
\newcommand{\lr}{\left(\lambda\mb{r}\right)}
\newcommand{\wt}{\widetilde}
\newcommand{\ddfrac}[2]{\frac{\mathrm{d}{#1}}{\mathrm{d}{#2}}}

\newcommand{\intx}{\int d^3x\ }
\newcommand{\intxp}{\int d^3x^{\prime}\ }
\newcommand{\intl}{\int_0^1 d\lambda\ }
\newcommand{\Dfrac}[2]{\frac{\partial{#1}}{\partial{#2}}}
\newcommand{\Pfrac}[2]{\frac{\delta{#1}}{\delta{#2}}}

\DeclareMathAlphabet\mbc{OMS}{cmsy}{b}{n}

\usepackage{etoolbox}
\preto\subequations{\ifhmode\unskip\fi} 

\usepackage[usenames,dvipsnames]{xcolor}
\usepackage[colorlinks=true,allcolors=black]{hyperref}
\newcommand{\bN}{\boldsymbol{\nabla}}

\begin{document}
	

\title{Avoiding gauge ambiguities in cavity quantum electrodynamics}

\author{Dominic M. Rouse}
\email{dmr9@st-andrews.ac.uk}
\affiliation{SUPA, School of Physics and Astronomy, University of St Andrews, St Andrews, KY16 9SS, UK}

\author{Brendon W. Lovett}
\affiliation{SUPA, School of Physics and Astronomy, University of St Andrews, St Andrews, KY16 9SS, UK}

\author{Erik M. Gauger}
\affiliation{SUPA, Institute of Photonics and Quantum Sciences, Heriot-Watt University, Edinburgh, EH14 4AS, UK}

\author{Niclas Westerberg}
\email{Niclas.Westerberg@glasgow.ac.uk}
\affiliation{School of Physics and Astronomy, University of Glasgow, Glasgow G12 8QQ,
United Kingdom}
\affiliation{SUPA, Institute of Photonics and Quantum Sciences, Heriot-Watt University, Edinburgh, EH14 4AS, UK}

\date{\today}%

\begin{abstract}
Systems of interacting charges and fields are ubiquitous in physics. Recently, it has been shown that Hamiltonians derived using different gauges can yield different physical results when matter degrees of freedom are truncated to a few low-lying energy eigenstates. This effect is particularly prominent in the ultra-strong coupling regime.  Such ambiguities arise because transformations reshuffle the partition between light and matter degrees of freedom and so level truncation is a gauge dependent approximation. To avoid this gauge ambiguity, we redefine the electromagnetic fields in terms of potentials for which the resulting canonical momenta and Hamiltonian are explicitly unchanged by the gauge choice of this theory. Instead the light/matter partition is assigned by the intuitive choice of separating an electric field between displacement and polarisation contributions. This approach is an attractive choice in typical cavity quantum electrodynamics situations.
\end{abstract}
	
\maketitle
\section{Introduction}
The 
gauge invariance of quantum electrodynamics (QED) is fundamental to the theory and can be used to greatly simplify calculations \cite{fiutak1963,babiker1983derivation,jackson2002lorenz,kok2010introduction,mahan2013many,rousseau2017quantum,andrews2018perspective,stokes2019gauge}. Of course, gauge invariance implies that physical observables are the same in all gauges despite superficial differences in the mathematics. However, it has recently been shown that the invariance is lost in the strong light/matter coupling regime if the matter degrees of freedom are treated as quantum systems with a fixed number of energy levels \cite{de2018breakdown,de2018cavity,stokes2018master,vukics2018gauge,rousseau2019reply,di2019resolution,stokes2019gauge}, including the commonly used two-level truncation (2LT). At the origin of this is the role of gauge transformations (GTs) in deciding the partition between the light and matter degrees of freedom, even if the primary role of gauge freedom is to enforce Gauss's law. Despite its long history \cite{goppert1931,power1959,yang1976,milonni1984,willis1987}, this has led to new questions about which gauge most accurately describes the physics.

\begin{figure}
\includegraphics[width=0.5\textwidth]{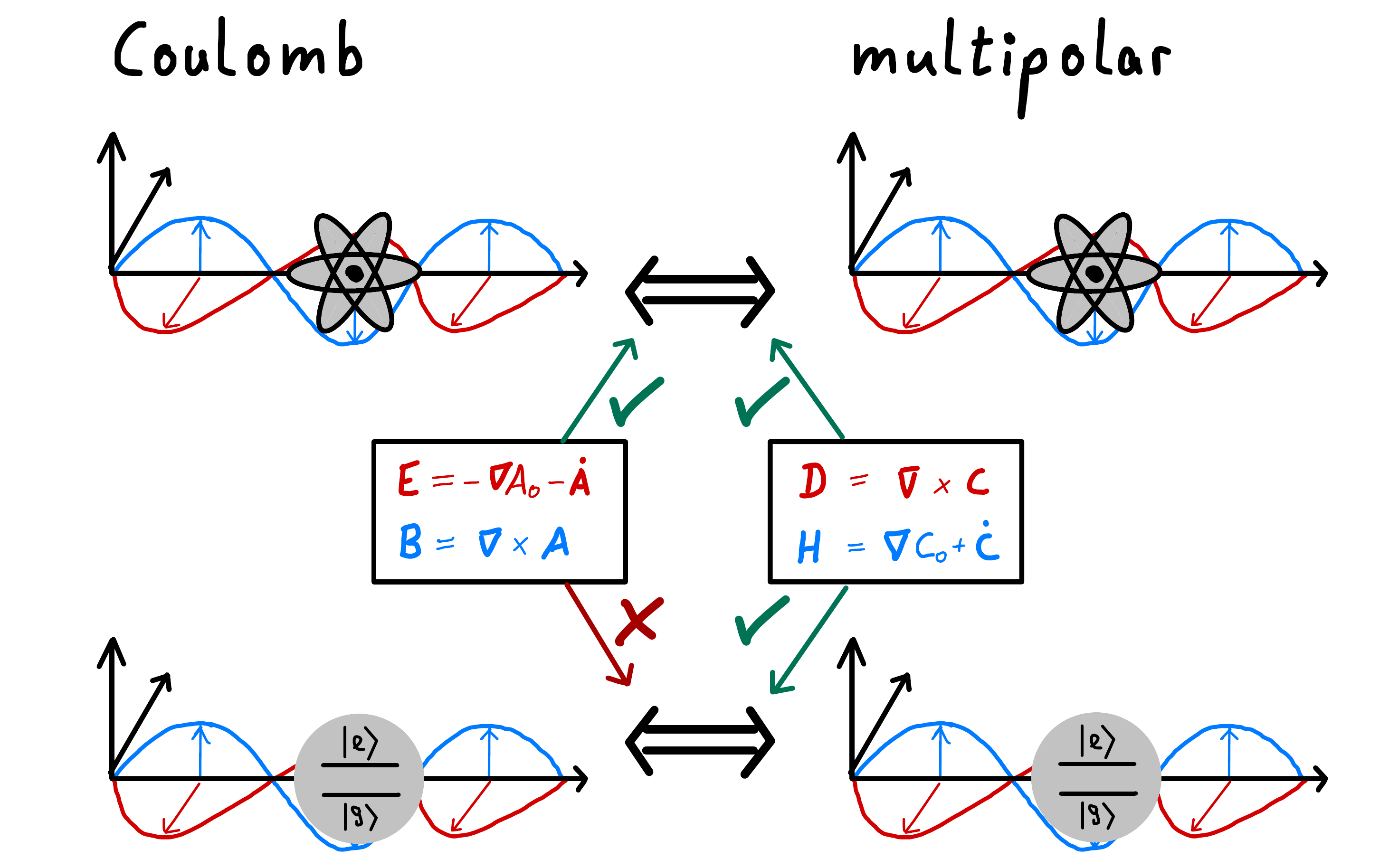}
\caption{A schematic of the electromagnetic potentials in the conventional and new approaches. On the left, the electromagnetic fields are parametrized using conventional $\mb{A}$- and $A_0$-fields, respectively, whereas on the right, the $\mb{C}$- and $C_0$-fields are used. The various relations and equivalences between the gauges are also shown. In particular note that gauge ambiguities manifest when the matter levels are truncated using the $\mb{A}$-fields.}\label{fig:sketch}
\vspace{-10pt}
\end{figure}	

Two common gauge choices of non-relativistic QED~\cite{de2018breakdown,de2018cavity,di2019resolution} are the Coulomb gauge, which has the advantage of describing photons as purely transverse radiation modes, and the multipolar gauge, which is most useful when the leading order (dipole) terms are dominant in a multipole expansion of the fields \cite{cohen1997photons,rokaj2018light,schafer2019relevance}. Interestingly, within the 2LT the multipolar gauge is usually found to agree more closely with exact, gauge invariant calculations than the Coulomb gauge \cite{de2018breakdown,de2018cavity}. This has been attributed to the fact that in the Coulomb gauge the light/matter interaction strength scales with the transition frequency between the relevant matter levels, while in the multipolar gauge the coupling instead scales with the energy of the radiation mode. Therefore, transitions between well-separated matter levels can be non-negligible in the Coulomb gauge \cite{de2018breakdown,de2018cavity}. Further, Ref.~\cite{di2019resolution} suggested that the 2LT in the Coulomb gauge converts the local potential into a non-local one, which no longer only depends on position but now also on the gauge-dependent canonical momentum. This is further discussed in Refs.~\cite{nori2019,garziano2020gauge,Taylor2020gauge} for a variety of physical settings.
	
The implication of these results is that the multipolar gauge is usually more accurate when the matter system is quantised and truncated to two levels. We should stress, however, that all gauges are equivalent and yield the same results if no approximations are made --- a situation in which the so-called Power-Zienau-Woolley Hamiltonian is appropriate (found in for instance Ref.~\cite{babiker1983derivation}). Nonetheless, after approximations are made (such as a 2LT or a Born-Oppenheimer approximation) separate gauges can yield differing results, since energy levels have different meanings in different gauges. With regards to this, it has been emphasised \cite{stokes2018master,stokes2019gauge} that there exist a continuum of possible GTs, each with a unique light/matter partition and therefore also 2LT; depending on the physical setting, gauges other than the common choices can offer more accurate 2LTs~\cite{stokes2018master,stokes2019gauge,roth2019optimal}. Recently, it has also been reported that time-dependent light/matter couplings can lead to gauge ambiguities \cite{stokes2019ultrastrong}.
	
In this Article, we first review the conventional approach in Section~\ref{sec:conven}, after which we reformulate QED such that gauge ambiguities do not manifest, see Fig.~\ref{fig:sketch}, in Section~\ref{sec:new}. We thus separate gauge issues from the choice of light/matter partition, for which we now offer an alternative interpretation. Our reformulation builds on previous work on the dual representation of QED \cite{bliokh2013dual,baker1994classical,bisht2008revisiting,chiao2004effective,hillery1984quantization,drummond_hillery_2014}. We then show that the dual representation recovers the multipolar-gauge Hamiltonian of the conventional theory when the light/matter partition is chosen appropriately. We also provide a physical explanation as to why this choice is optimal for systems typical of QED, e.g. dipoles in a cavity. Additionally, in Section~\ref{sec:2LT}, we numerically compare the accuracy of 2LTs in different light/matter partitions for the example of dipoles in a cavity.  The results are discussed in Section~\ref{sec:discuss}. We provide further details, extensions to the model, and derivations in the Supplementary Material.

\section{Conventional approach}\label{sec:conven}
We first outline the conventional approach, and how gauge ambiguities arise in it (see Supplementary Material~1 for full mathematical details). Let us consider a generic system of charges $q_{\mu}$ at positions $\mb{r}_{\mu}$ described by a charge density $\rho$ and current density $\mb{J}$. Their dynamics are governed by the Maxwell equations and Lorentz force:
\begin{subequations}\label{eq:ML}
	\begin{align}
	&\dive \mb{B}\x=0,\label{eq:M1}\\ 
	&\curl \mb{E}\x=-\dot{\mb{B}}\x,\label{eq:M2}\\
	&\dive \mb{E}\x=\rho\x/\varepsilon_0,\label{eq:M3}\\
	&\curl \mb{B}\x=\mu_0\mb{J}\x+\varepsilon_0\mu_0\dot{\mb{E}}\x,\label{eq:M4}\\
	&m_{\mu}\ddot{\mb{r}}_{\mu}=q_{\mu}\left[\mb{E}(\mb{r}_{\mu})+\dot{\mb{r}}_{\mu}\times\mb{B}(\mb{r}_{\mu})\right].\label{eq:M5}
	\end{align}
\end{subequations}

Conventionally, the electric and magnetic fields are parametrized in terms of vector and scalar potentials $\mb{A}$ and $A_0$ as $\mb{E}= -\bN A_0-\dot{\mb{A}}$ and $\mb{B} = \curl\mb{A}$ respectively, leading immediately to Faraday's law and sourceless magnetic fields~\cite{jackson}. The remaining equations are derived by minimizing the action of the \textit{minimal-coupling Lagrangian} \cite{babiker1983derivation,andrews2018perspective, stokes2019gauge,kok2010introduction,mahan2013many}
\begin{equation}
L_m=\sum_{\mu}\frac{1}{2}m_{\mu}\dot{\mb{r}}^2_{\mu}+\intx\mathcal{L}_m(\mb{x}),\label{eq:Lminimal}
\end{equation}
with the Lagrange density
\begin{align}
\mathcal{L}_m(\mb{x}) = \frac{\varepsilon_0}{2}\left[\mb{E}^2 - c^2\mb{B}^2\right]+\left[\mb{J}\cdot\mb{A}\x-\rho A_0\right],\label{eq:LDminimal}
\end{align}
where the mechanical degrees of freedom are $A_0$, $\mb{A}$ and $\mb{r}_\mu$, respectively. The physical fields are unchanged by the introduction of a scalar field
\begin{equation}
\chi\x=\intxp\tilde{\chi}\xx,
\end{equation}
so long as
\begin{subequations}
\begin{eqnarray}
&&A_0\p\x=A_0\x+\dot{\chi}\x,\\
&&\mb{A}\p\x=\mb{A}\x-\bN\chi\x,\label{eq:gaugetrans}
\end{eqnarray}
\end{subequations}
where a primed variable indicates one transformed by the $\chi$-field. Under this transformation the Lagrangian $L\p_m$ has a modified Lagrange density given by
\begin{align}
{\mathcal{L}_m\p\x} = \mathcal{L}_m\x &- \mb{J}\x\cdot\bN\chi\x \nonumber\\
&-\rho\x\intxp\dot{\mb{A}}\xp\cdot\Dfrac{\tilde{\chi}\xx}{\mb{A}\xp}.
\end{align}
The canonical momenta of this arbitrary-gauge Lagrangian, $\mb{p}_{\mu}\p=\partial L\p_m/\partial\dot{\mb{r}}_{\mu}$ and $\mb{\Pi}\p=\delta\mathcal{L}\p_m/\delta \dot{\mb{A}}$, can be found as \cite{babiker1983derivation}
\begin{subequations}
\begin{eqnarray}
&&\mb{p}_{\mu}\p=m_{\mu}\dot{\mb{r}}_{\mu}+q_{\mu}\mb{A}(\mb{r}_{\mu})-q_{\mu}\bN\chi(\mb{r}_{\mu}),\\
&&\mb{\Pi}\p\x=-\varepsilon_0\mb{E}\x-\bm{\phi}\p\x
\end{eqnarray}
\end{subequations}
where 
\begin{equation}
\bm{\phi}\p\x=\intxp\rho\xp\Dfrac{\tilde{\chi}\xx}{\mb{A}\x}.
\end{equation} 

Importantly, $\mb{p}_{\mu}\p$ and $\mb{\Pi}\p$ are explicitly gauge dependent and so correspond to different canonical momenta in every gauge \cite{babiker1983derivation,stokes2019gauge}. After eliminating $A_0$ using the continuity equation, the arbitrary-gauge Hamiltonian is found as \cite{babiker1983derivation}
\begin{align}\label{eq:HOld}
H\p=\sum_{\mu}&\frac{1}{2m_{\mu}}\bigg[\mb{p}\p_{\mu} - q_{\mu}\mb{A}(\mb{r}_{\mu})+q_{\mu}\bN\chi(\mb{r}_{\mu})\bigg]^2 \\
&+\intx\left(\frac{1}{2\varepsilon_0}\left[\mb{\Pi}\p\x+\bm{\phi}\p\x\right]^2 + \frac{\mb{B}^2\x}{2\mu_0}\right).\nonumber
\end{align} 

Gauge ambiguities can occur, particularly in the strong light/matter coupling regime, when approximations to the Hamiltonian are introduced. A prominent example of this is expressing the matter Hamiltonian using a truncated number of energy levels; this approximation has different meanings in each gauge. When quantizing, the gauge dependent classical momentum $\mb{p}\p_\mu$ is promoted to its quantum counterpart $\mb{\widehat{p}\p_\mu}$ (along with the position $\mb{r}_\mu\rightarrow\mb{\hat{r}}_\mu$). The truncation to $N+1$ discrete energy levels follows next for each charge:
\begin{equation}
{\mb{\widehat{T}}\p_{\mu}\equiv\frac{\mb{\widehat{p}}^{\prime 2}_{\mu}}{2m_{\mu}}+\widehat{U}_{\mathrm{ext}}\left(\mb{\hat{r}}_\mu\right)}\to\sum_{n=0}^N\epsilon_{n,\mu}\proj{\epsilon_{n,\mu}\p}{\epsilon_{n,\mu}\p},\label{eq:pquant}
\end{equation}
where $\widehat{U}_{\mathrm{ext}}$ is the external electrostatic interaction binding the charges. Each `matter' eigenstate in Eqn.~\eqref{eq:pquant} refers to a different physical system in each gauge and so truncation means losing different information. Formally, only when $N\to\infty$ do all observables agree in different gauges, though for weak light/matter coupling a low-level truncation is usually sufficient for good agreement.

\section{New approach}\label{sec:new}
The canonical momenta in the theory outlined above inherit their gauge-dependency from the minimal-coupling Lagrangian [Eqn.~\eqref{eq:Lminimal}], as the vector and scalar potentials are only defined up to the scalar function $\chi$. To remove gauge ambiguities, we will therefore derive a theory which is described by a Lagrangian depending only on the physical fields. 
	
The total charge and current densities of any system can be partitioned into \textit{bound} and \textit{free} contributions as $\rho=\rho_b+\rho_f$ and $\mb{J}=\mb{J}_b+\mb{J}_f$ \cite{jackson}. This naturally allows one to distinguish two contributions to the electric and magnetic fields: $\mb{E}=(\mb{D}-\mb{P})/\varepsilon_0$ and $\mb{B}=\mu_0(\mb{H}+\mb{M})$ with $\mb{P}$ and $\mb{M}$ being the polarisation and magnetisation fields. Our aim now is to parametrize the displacement and magnetic fields $\mb{D}$ and $\mb{H}$ using a dual vector potential $\mb{C}$ and scalar potential $C_0$ such that
\begin{subequations}\label{eq:EMfields2}
\begin{align}
\mb{D}\x&=\curl\mb{C}\x,\label{eq:EMfields2a}\\
\mb{H}\x&=\bN C_0\x+\dot{\mb{C}}\x.\label{eq:EMfields2b}
\end{align}
\end{subequations}
This is the crucial point of this Article, and as we will show, it avoids gauge ambiguities in the formulation of cavity QED. The parametrization in terms of $\mb{C}$-fields relies on the absence of free currents $\mb{J}_f$, a common cavity QED setting~\cite{de2018breakdown,de2018cavity, stokes2018master,di2019resolution,stokes2019gauge}. Other examples of defining the physical fields in this way can be found in \cite{bliokh2013dual,baker1994classical,bisht2008revisiting,chiao2004effective,hillery1984quantization,drummond_hillery_2014}, although here we extend the formulation to include the magnetization field and therefore move beyond the standard electric dipole approximation.

The polarization field $\mb{P}$ and the magnetisation field $\mb{M}$ are sourced by the bound charge and currents, respectively:
\begin{subequations} \label{eq:PMdef}
\begin{align}
\dive\mb{P}\xr &= -\rho_b\xr,\label{eq:diveP}\\
\curl\mb{M}\xr &= \mb{J}_b\xr- \dot{\mb{P}}\xr.\label{eq:Jbgeneral}
\end{align}
\end{subequations}
We also note that Maxwell's equations Eqns.~\eqref{eq:M3} and \eqref{eq:M4} become:
\begin{subequations}\label{eq:MLb}
\begin{align}
\dive \mb{D}\x &= \rho_f\x,\label{eq:Mb3}\\
\curl \mb{H}\x &= \mb{J}_f\x+\dot{\mb{D}}\x,\label{eq:Mb4}
\end{align}
\end{subequations}
when written in terms of the displacement field $\mb{D}$ and magnetic field $\mb{H}$. Note that interestingly, within this formalism Eqns.~\eqref{eq:M1}-\eqref{eq:M2} and Eqns.~\eqref{eq:M3}-\eqref{eq:M4} switch roles, as Eqns.~\eqref{eq:M1}-\eqref{eq:M2} are dynamical equations for the $\mb{C}$-field with Eqns.~\eqref{eq:M3}-\eqref{eq:M4} serving as the Bianchi identity, whereas the opposite is true for the $\mb{A}$-field.

We now specify a system to illustrate the theory, and for simplicity we will choose a single dipole formed of an electron at position $\mb{r}$ and a hole at the origin. The bound charge density and current of this dipole are described by $\rho_b\xr = -e\delta(\mb{x}-\mb{r})+e\delta(\mb{x})$ and $\mb{J}_b\xr = -e\dot{\mb{r}}\delta(\mb{x}-\mb{r})$, respectively. The theory is easily extended to more dipoles, and in Supplementary Material~2 we add a background ionic lattice which allows for phonon-mediated processes within the system. There are no free charges or currents ($\rho_f=\mb{J}_f=0$) and so a symmetry emerges when comparing Eqns.~\eqref{eq:MLb} to Maxwell's equations Eqns.~\eqref{eq:M1} and \eqref{eq:M2} \cite{jackson,cohen1997photons,bliokh2013dual,baker1994classical,bisht2008revisiting,chiao2004effective,hillery1984quantization,drummond_hillery_2014}. We will exploit this symmetry to parametrize the displacement and magnetic fields according to Eqns.~\eqref{eq:EMfields2}.

The restrictions on $\mb{P}$ and $\mb{M}$ given by Eqns.~\eqref{eq:PMdef} produce the correct bound charge density and current if \cite{babiker1983derivation,stokes2019gauge,stokes2019ultrastrong}
\begin{equation}
\mb{P}\xr=-e\intl\mb{r} \delta(\mb{x}-\lambda\mb{r}),
\end{equation}
and 
\begin{equation}
\mb{M}\xr=-\dot{\mb{r}}\times\bm{\theta}(\mb{x},\mb{r}),
\end{equation}
where 
\begin{equation}
\bm{\theta}(\mb{x},\mb{r})=-e\intl\lambda\mb{r}\delta(\mb{x}-\lambda\mb{r}).
\end{equation} 
However, these are not unique and Eqns.~\eqref{eq:PMdef} are also satisfied by $\mb{P}\to\wt{\mb{P}}=\mb{P}+\wt{\mb{P}}_V$ and $\mb{M}\to\wt{\mb{M}}=\mb{M}-\wt{\mb{M}}_V$ where
\begin{subequations}\label{eq:PMtrans}
\begin{align}
&\wt{\mb{P}}_V\xr=\curl\mb{V}\xr,\\
&\wt{\mb{M}}_V\xr=\dot{\mb{V}}\xr+\bN V_0\xr,
\end{align}
\end{subequations}
for general fields $\mb{V}$ and $V_0$ and quantities dependent on these are denoted with a tilde. Such a transformation does not change the physics, but alters the light/matter partition. We emphasize that, in contrast, for the conventional $\mb{A}$-field theory the light/matter partition is encompassed by gauge freedom.

\begin{figure*}\centering
\includegraphics[width=0.8\textwidth]{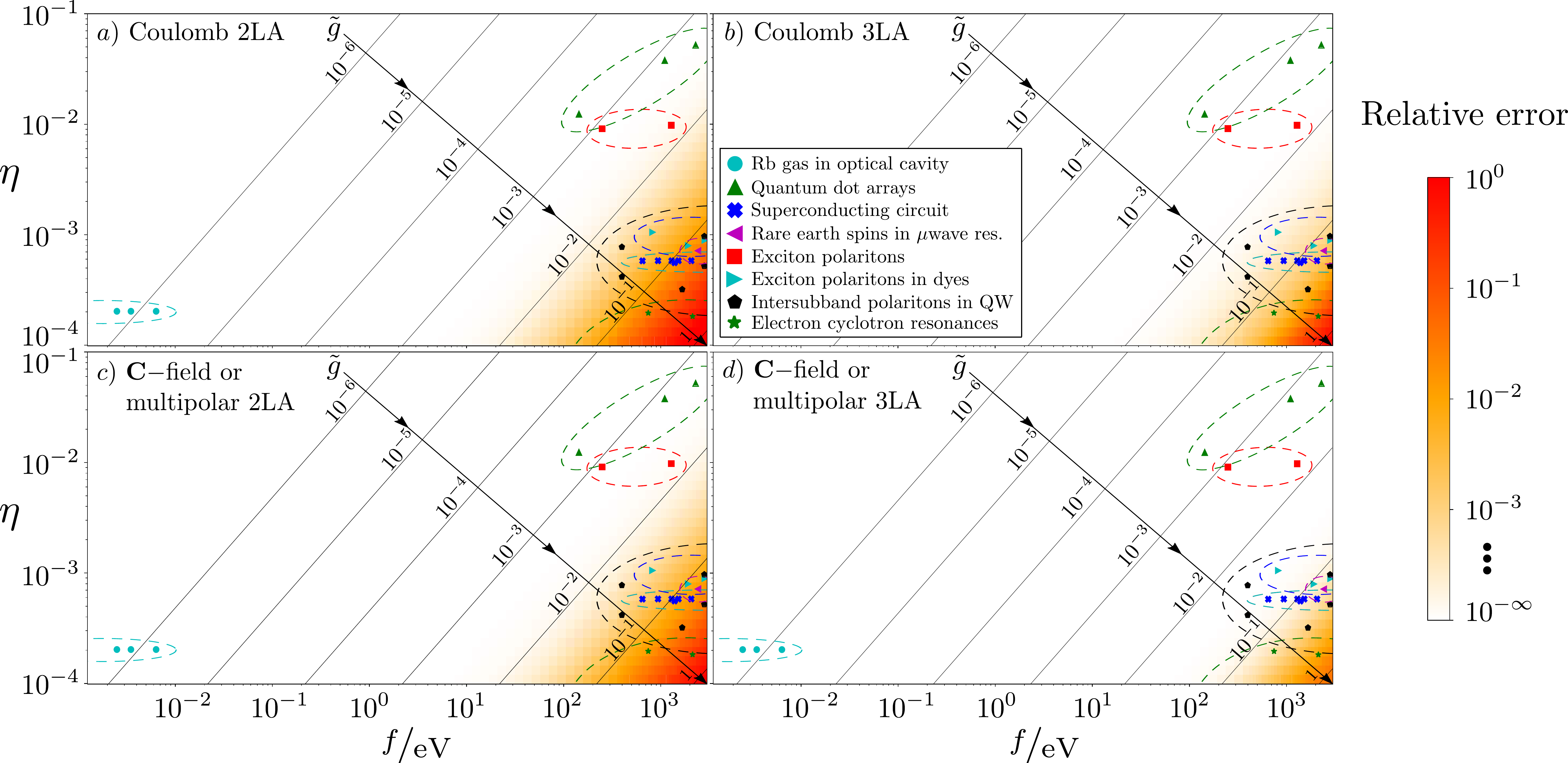}
\caption{The relative error in calculating the lowest energy spacing of the full Coulomb gauge and $\mb{C}$-field Hamiltonians for an infinite square well potential. The axes are $\eta=(1/2\pi)x_{10}\nu = x_{10}/\lambda_{\mathrm{rad}}$, where $x_{10}=\left<\epsilon_0\p\right|\mb{x}\left|\epsilon_{1}\p\right>$ is the approximate dipole size. The electric dipole approximation is satisfied when $\eta\ll1$. To vary $\eta$, we vary $\nu$ whereas $x_{10}$ is constrained such that the first dipole transition is resonant with the radiation mode ($\epsilon_1-\epsilon_0=\nu$), which in practice results in tuning the length of the well. Along the horizontal axis, we plot the magnitude of the vector potential $f$. Importantly, the physically relevant second axis $\tilde{g} = g^C_{10}/\omega_{10}=g^m_{10}/\nu$ is plotted on the diagonal where $g^i_{10}$ is the transition strength between the two lowest lying states in gauge $i$. Here $\tilde{g}>1$ indicates ultra-strong coupling. We also show approximate regions where different types of QED experiments sit with respect to $\eta$ and $\tilde{g}$ in the plot, with markers indicating individual experiments given in Table~SM1 of Supplementary Material~5. References for the experimental data: \textit{Rb gas in optical cavity} \cite{suleymanzade2019tunable,colombe2007strong,thompson2013coupling,tiecke2014nanophotonic}; \textit{Quantum dot arrays} \cite{gerard1998enhanced,gerard1996quantum,gerard1998inas,gerard1998inas,gayral1999high,moreau2001single}; \textit{Superconducting circuits} \cite{johansson2006vacuum,niemczyk2010circuit,forn2010observation,baust2016ultrastrong,yoshihara2017superconducting}; \textit{Rare earth spins in $\mu$wave resonator} \cite{weisbuch1992observation,bloch1998giant}; \textit{Exciton polaritons} \cite{weisbuch1992observation,bloch1998giant}; \textit{Exciton polaritons in dyes} \cite{bellessa2004strong,wei2013adjustable,gambino2014exploring,kena2013ultrastrongly}; \textit{Intersubband polaritons in quantum wells} \cite{dupont2003vacuum,dupont2007giant,todorov2010ultrastrong,delteil2012charge,askenazi2014ultra}; \textit{Electron cyclotron resonances} \cite{muravev2011observation,scalari2012ultrastrong,maissen2014ultrastrong,bayer2017terahertz}.} \label{fig:2}
\end{figure*}

All that remains to complete the theory is to write a Lagrangian that reproduces the remaining Maxwell equations [Eqns.~\eqref{eq:M1} and \eqref{eq:M2}], and Lorentz force equation [Eqn.~\eqref{eq:M5}], when minimised with respect to the mechanical degrees of freedom $C_0$, $\mb{C}$ and $\mb{r}$ respectively. We find that the required Lagrangian is
\begin{align}\label{eq:Lnew}
L&=\frac{1}{2}m\dot{\mb{r}}^2+\intx \frac{\varepsilon_0}{2}\left[c^2\mb{B}^2\x-\mb{E}^2\x\right]\\
&= \frac{1}{2}m\dot{\mb{r}}^2+\intx \frac{\varepsilon_0}{2}\bigg[c^2\left(\dot{\mb{C}}+\boldsymbol{\nabla}C_0 + \mb{M}\right)^2\nonumber\\
&\hspace{4.8cm}-\left(\boldsymbol{\nabla}\times\mb{C}-\mb{P}\right)^2\bigg],\nonumber
\end{align}
which for $\mb{M} \rightarrow \mb{0}$ agrees with Refs.~\cite{drummond2006dual,drummond_hillery_2014} and we prove in Supplementary Material~3 that this Lagrangian satisfies all the necessary equations of motion. 

Eqns.~\eqref{eq:EMfields2} are invariant under gauge transformations $C_0\to C_0+\dot{\xi}$ and $\mb{C}\to\mb{C}-\bN\xi$ for any arbitrary scalar field $\xi$, but importantly so is the Lagrangian in Eqn.~\eqref{eq:Lnew}. This is because the Lagrangian is written only in terms of the physical fields. Additionally, this means that the Lagrangian is invariant under the transformations in Eqns.~\eqref{eq:PMtrans}. It is also possible to verify this with a Lagrangian written in terms of the mechanical variables $C_0$, $\mb{C}$ and $\mb{r}$. For $\mb{B}$ and $\mb{E}$ to be invariant under this transformation, there must be an implicit change to the fields $\mb{C}$ and $C_0$, which we write explicitly as $\mb{D}\to\wt{\mb{D}}$ and $\mb{H}\to\wt{\mb{H}}$ where
\begin{align}
    &\wt{\mb{D}}\x\equiv\curl\wt{\mb{C}}\x=\mb{E}\x+\wt{\mb{P}}\xr,\\
    &\wt{\mb{H}}\x\equiv\dot{\wt{\mb{C}}}\x+\nabla\wt{C}_0\x=\mb{B}\x-\wt{\mb{M}}\xr.
\end{align}
This is a direct consequence of the transformation in Eqns.~\eqref{eq:PMtrans} changing the light/matter partition; a redistribution of the contributions of $\mb{D}\pp$ and $\mb{P}\pp$ to $\mb{E}\pp$, and likewise $\mb{H}$ and $\mb{M}$ to $\mb{B}$. Note that we have here introduced the Helmholtz decomposition of a vector $\mb{W}=\mb{W}\pll+\mb{W}\pp$ into parallel $\mb{W}\pll$ and perpendicular $\mb{W}\pp$ components, satisfying $\curl\mb{W\pll}=0$ and $\dive\mb{W}\pp=0$ respectively. 

The gauge-invariant Lagrangian in Eqn.~\eqref{eq:Lnew} leads to the crucial result that the canonical momenta in the new theory are also no longer gauge dependent, although they do depend on the light/matter partition through $\mb{V}$ and $V_0$. We find that the canonical momenta are
\begin{subequations}\label{eq:CM2}
\begin{align}
&\wt{\mb{p}}=\Dfrac{\wt{L}}{\dot{\mb{r}}}=m\dot{\mb{r}}-\bm{\Phi}_0\rr-\wt{\bm{\Phi}}\rr,\label{eq:CM2a}\\
&\wt{\mb{\Pi}}\x=\Pfrac{\wt{\mathcal{L}}}{\dot{\wt{\mb{C}}}\x}=\mb{B}\x,\label{eq:CM2b}
\end{align}
\end{subequations}
where $\bm{\Phi}_0\rr=\intx\bm{\theta}(\mb{x},\mb{r})\times\mb{B}\x$ and
\begin{equation}
\wt{\bm{\Phi}}\rr=\Dfrac{}{\dot{\mb{r}}}\intx\mb{B}\x \cdot\wt{\mb{M}}_V\xr.
\end{equation}
We derive Eqns.~\eqref{eq:CM2} in Supplementary Material~4. We see that the field canonical momentum is always the magnetic field whilst the matter canonical momentum is dependent on the light/matter partition. 

To derive the Hamiltonian, we must be able to invert Eqn.~\eqref{eq:CM2a} to write $\dot{\mb{r}}$ as a function of $\mb{p}$. This puts a constraint on the allowed $\mb{V}$ and $V_0$ fields in the transformations in Eqns.~\eqref{eq:PMtrans}. Here we assume that this constraint is met which results in $\wt{\bm{\Phi}}$ being independent of $\dot{\mb{r}}$, in which case we find that the Hamiltonian is
\begin{eqnarray}
\wt{H}=&&\frac{1}{2m}\left[\wt{\mb{p}}+\bm{\Phi}_0\rr+\wt{\bm{\Phi}}\rr\right]^2+\widehat{U}_{\mathrm{ext}}\label{eq:Hnew}\\
&&+\intx \left(\frac{\mb{B}^2\x}{2\mu_0} + \frac{1}{2\varepsilon_0}\left[\wt{\mb{D}}\pp\x-\wt{\mb{P}}\x\right]^2\right),\nonumber
\end{eqnarray}
where we have introduced an external potential $\widehat{U}_{\mathrm{ext}}$. Eqn.~\eqref{eq:Hnew} is derived explicitly in Supplementary Material~4 but follows the standard procedure. The gauge independence of Eqn.~\eqref{eq:Hnew} follows from the absence of magnetic monopoles, and as such the primary constraint $\bN\cdot\mb{B} = 0$ can be satisfied without altering the light/matter partition. Note that, differently to $\mb{A}$-field theory, Gauss's law can be enforced as an initial condition \cite{drummond_hillery_2014}. The constraint on inverting Eqn.~\eqref{eq:CM2a} manifests as an additional term in Eqn.~\eqref{eq:Hnew} with the form
\begin{equation}\label{eq:constraint}
    \left(1-\dot{\mb{r}}\cdot\Dfrac{}{\dot{\mb{r}}}\right)\intx\mb{B}\x\cdot\wt{\mb{M}}_V\xr.
\end{equation}
This term vanishes when Eqn.~\eqref{eq:CM2a} can be inverted, e.g. for $(\mb{V},V_0)=(\mb{0},0)$ and $(\mb{V},V_0)=(\frac{1}{2}\mb{r}\times\mb{x},0)$.

Before quantising the fields, we must choose a light/matter partition. In the conventional theory this requires a choice of gauge, however, the gauge choice here does not alter this partition. Instead, this freedom is encompassed in the choice of $\mb{P}\pp$ and $\mb{M}$. We show now that by choosing $\mb{V}=V_0=0$ we arrive at the usual multipolar gauge Hamiltonian of the conventional theory. This means that choosing $\mb{V}=V_0=0$ must result in the same light/matter partition as that in the multipolar gauge of the conventional theory. After making this choice we can now remove the tildes on the fields. We must then also choose a gauge in order for the quantisation procedure to be well-defined. This is because there are redundant variables in the Lagrangian, just as in a free $\mb{A}$-field theory. Here we pick the Coulomb-gauge analogue of $\bN\cdot\mb{C}=0$ and $C_0 = 0$, but we note that the gauge does not affect the light/matter partition nor the form of the Hamiltonian. We quantise the fields by enforcing $\left[\widehat{C}\pp_i\qx,\widehat{\Pi}_j\qxp\right]=i\delta\pp_{ij}(\mb{x}-\mb{x}\p)$, where
\begin{equation}
\mb{\widehat{C}}\pp\qx=\sum\kl \bm{\epsilon}\kl f\uk\left(\hat{\mathfrak{c}}\kl^{\dagger}\rme^{-i\mb{k}\cdot\mb{x}}+\hat{\mathfrak{c}}\kl\rme^{i\mb{k}\cdot\mb{x}}\right),\label{eq:C}
\end{equation}
and $f\uk=(2\nu\uk V)^{-{1/2}}$ is the coupling strength to the mode with frequency $\nu\uk=c|\mb{k}|$ in volume $V$, $\bm{\epsilon}\kl$ are polarisation vectors orthonormal to $\mb{k}$, and ${\hat{\mathfrak{c}}}\kl$ (${\hat{\mathfrak{c}}}\kl^{\dagger}$) is the photon annihilation (creation) operator for the $\mb{C}$-field. We note that for different choices of $\mb{V}$ and $V_0$, the ladder operators in $\widehat{{\mb{C}}}\pp$ describe different bosons. After the matter parts of the Hamiltonian are also quantized and expanded into eigenstates (truncated to $N+1$ levels) the Hamiltonian can be written as 
\begin{align}
\widehat{H} = &\sum_{n=0}^N \epsilon_{n}\proj{\epsilon_{n}}{\epsilon_{n}}+\sum\kl \nu_{\mb{k}} \hat{\mathfrak{c}}^{\dagger}\kl \hat{\mathfrak{c}}\kl\label{eq:HnewQ}\\
&-\frac{1}{\varepsilon_0}\mb{\widehat{d}}\cdot\left[\curl\mb{\widehat{C}}\pp(\mb{0})\right] +\frac{1}{\varepsilon_0}\sum\kl f_{\mb{k}}^2\nu_{\mb{k}}\left(\mb{\widehat{d}}\cdot\bm{\epsilon\kl}\right)^2,\nonumber
\end{align}
where $\widehat{\mb{d}}=-e\hat{\mb{r}}$. To arrive at Eqn.~\eqref{eq:HnewQ}, we make the electric dipole approximation (EDA) ($\mb{k}\cdot\mb{r}\ll1$) which allows us to evaluate the fields at the origin, set $\mb{P}\simeq-e\mb{r}\delta(\mb{x})$ and ignore the smaller magnetisation interactions governed through $\bm{\Phi}_0$. The quantization process is analogous to the $\mb{A}$-field theory, which is given in detail in Supplementary Material~3. We are now free to choose polarisation vectors in such a way that the polarisation of the physical fields as computed using $\mb{C}$ and $\mb{A}$-fields overlap. It then follows that the $\mb{C}$-field Hamiltonian in Eqn.~\eqref{eq:HnewQ} has the same mathematical form as the multipolar gauge Hamiltonian for the $\mb{A}$-field, which is a reflection of the light/matter partitions being identical. In Table~\ref{tab:comparison} we highlight the differences between the $\mb{A}$- and $\mb{C}$-field approaches.
\begin{center}
	\begin{table}[h]
		\caption{\label{tab:comparison}
			Comparison of $\mb{A}$-field and $\mb{C}$-field representations. Gauge dependent parameters are denoted with a prime. $C$/$m$ (``$C$/$m$'') denotes the Coulomb/multipolar gauge (-analogues) in the $\mb{A}$-field ($\mb{C}$-field) representation respectively, however the choice of gauge is inconsequential for predictions of the $\mb{C}$-theory. We define $\bm{\Phi}_{\mb{D}}\rr=\intx\bm{\theta}(\mb{x},\mb{r})\times\mb{D}\x$.}	
		\begin{ruledtabular}
		\begin{tabular}{cc}
			$\mb{A}$-field approach & $\mb{C}$-field approach \\
			\colrule 
			$\mb{B}=\curl\mb{A}$ & $\mb{H}=\bN C_0+\dot{\mb{C}}$ \\ 
			$\mb{E}=-\bN A_0-\dot{\mb{A}}$ & $\mb{D}=\curl\mb{C}$ \\ 
			$\mb{A}\p= \begin{cases} \mb{A}\pp & C \\ \bm{\Phi/e} & m \end{cases}$& $\mb{C}\p= \begin{cases} \mb{C}\pp & ``C" \\ \bm{\Phi}_{\mb{D}}/e & ``m" \end{cases}$\\
			$\bm{\Pi}\p= \begin{cases} -\varepsilon_0\mb{E}\pp & C \\ -\mb{D} & m \end{cases}$& $\wt{\bm{\Pi}}= \mb{B}$\\
			$\mb{p}\p=\begin{cases} m\dot{\mb{r}}+e\mb{A}\pp  & C \\ m\dot{\mb{r}}-\bm{\Phi} & m \end{cases}$ & $\wt{\mb{p}}=m\dot{\mb{r}}-\bm{\Phi}_0-\wt{\bm{\Phi}}$\\
		\end{tabular}
		\end{ruledtabular}
	\end{table}
\end{center}

\section{Accuracy of two-level truncations}\label{sec:2LT}
We now turn to the question of whether a 2LT for the matter system is possible. In Figure~\ref{fig:2} we display the accuracy of the $\mb{C}$-field Hamiltonian in an arbitrary gauge, with $\mb{P}^\perp$ and $\mb{M}$ chosen such that the light/matter partition is equivalent to the $\mb{A}$-field in the multipolar gauge, along with the conventional $\mb{A}$-field in the Coulomb gauge. In both cases we truncate to two or three dipole levels, which we give details on shortly. Here, we use only a single radiation mode that is resonant with the transition between the two lowest dipole levels, and $\widehat{U}_{\mathrm{ext}}$ is an infinite square well potential whose anharmonicity makes it amenable to few-level expansion. In the strong coupling limit, $\tilde{g}\to 1$, both truncated Hamiltonians become inaccurate, importantly for different reasons. As discussed in Ref.~\cite{de2018breakdown}, we expect a theory that limits the coupling between states far separated in energy space to most accurately model the physics, such as our $\mb{C}$-field Hamiltonian with $(\mb{V},V_0)=(\mb{0},0)$ or, equivalently, a multipolar $\mb{A}$-field Hamiltonian. In such a theory, we expect the dynamics to be limited to a manifold containing few states, and so accuracy is much improved by going from two to three levels. In contrast, the Coulomb gauge couples many energy states, and should thus be inaccurate when truncated to two, or three, levels. 

We here appeal to discussion of the physics of the situation as the most natural way of determining a sensible light/matter partition. Importantly, there are two length-scales of the problem: the size of the dipole $L$ and the wavelength of the light $\lambda$. First, for polarisation fields to be well-approximated by a dipole moment at the origin, i.e. $\mb{P}\propto \widehat{\mb{d}}\delta(\mb{x})$, we must have $\lambda \gg L$. Polarisation fields can, of course, be used nonetheless, but at a computational cost. Second, it is easy to see that the transition dipole moment $\mb{d}_{n,m}$ scales with the size of the dipole $L$, as $|\langle \widehat{\mb{d}}\rangle| \propto \left|\left\langle \hat{\mb{r}}\right\rangle\right| \propto L$. Similarly, the momentum expectation value must scale as $1/L$, from unit considerations. Thus for a small dipole where $L \ll \lambda$, the momentum matrix elements $\mb{p}_{n,m}$ become large, whereas the dipole matrix elements $\mb{d}_{n,m}$ are small. This therefore necessarily limits the coupling between well-separated energy states when relying on a dipolar coupling, allowing the dynamics to take place in a small energy manifold. For a large dipole, the situation is reversed, and we should note that the polarisation field becomes computationally more intensive to use in the same limit (i.e. higher order multipolar modes must be accounted for). This suggests a physical origin to the success of the $\mb{C}$-field/multipolar gauge $\mb{A}$-field Hamiltonians. Indeed, we only see limited improvement by going from two to three levels for the Coulomb gauge. This is further discussed in Supplementary Material~6 where we repeat the calculation with a non-resonant cavity mode, shown in Figure~SM1. We should finally also note that in both cases we must keep counter-rotating terms, as they contribute significantly in the strong-coupling regime \cite{feranchuk2020exact}.

We now give details on the numerics performed in Figure~\ref{fig:2}. The  Hamiltonians have Hilbert spaces $\mathcal{H}_m\otimes\mathcal{H}_p$, where $\mathcal{H}_m$ ($\mathcal{H}_p$) is the $N_m$ ($N_p$) dimensional Hilbert space of the matter (single mode photon field). Written in matrix form the Coulomb gauge Hamiltonian of the conventional QED formulation is
\begin{align}
    \mbc{H}_{\mathrm{Cb}}&=\mbc{H}_m\otimes \mb{I}_{N_p}+\frac{e }{m}\mb{p}\otimes\mb{A}\nonumber\\
    &+\frac{e^2}{2m}\mb{I}_{N_m}\otimes\mb{A}^2+\nu \mb{I}_{N_m}\otimes \mb{a}^{\dagger}\cdot\mb{a},\label{eq:HCoulombmatrix}
\end{align}
where $e$ and $m$ are the electron charge and mass, $\nu$ is the energy of the photon mode and $\mb{I}_{d}$ is the identity operator of dimension $d$. Eqn.~\eqref{eq:HCoulombmatrix} is derived in Supplementary Material~1. The vector potential is $\mb{A}=f(\mb{a}+\mb{a}^{\dagger})$ where $\mb{a}$ is the annihilation operator matrix of dimension $N_p$ and $f$ is the field amplitude. Note that throughout this example we assume that the dipole aligns with the polarisation of the field mode. The matter energy levels are contained within
\begin{equation}
    \mbc{H}_m=\sum_{n=1}^{N_m}\epsilon_{n}\proj{\epsilon_{n}}{\epsilon_{n}},
\end{equation}
where $\ket{\epsilon_n}$ and $\epsilon_n$ are the eigenstate and eigenenergy solutions to the Schr\"odinger equation, and $n$ is an integer. In Figure~\ref{fig:2} we use the one dimensional infinite square well potential which is zero within $0\le x\le L$ and infinite outside this range. This leads to the well known eigenenergies and position space wavefunctions
\begin{align}
    &\epsilon_n=\frac{\pi^2 n^2}{2mL^2},\\
    &\psi_n(x)=\braket{x|\epsilon_n}=\sqrt{\frac{2}{L}}\sin\left(\frac{n\pi x}{L}\right).
\end{align}
Finally, the momentum matrix is $\mb{p}=\sum_{n,m=1}^{N_m}p_{n,m}\proj{\epsilon_n}{\epsilon_m}$ where the matrix elements are
\begin{equation}
    p_{n,m}=\begin{cases} \frac{4\hbar}{iL}\frac{nm}{n^2-m^2} & n+m\text{ odd} \\ 0 & n+m\text{ even}, \end{cases}
\end{equation}
 with $n,m$ integer. For a detailed reference on the infinite square well see Ref.~\cite{prentis2014matrix}. 

For a single photon mode the $\mb{C}$-field Hamiltonian with $(\mb{V},V_0)=(\mb{0},0)$ (and equivalently multipolar of the conventional QED formulation) is
\begin{align}
\mbc{H}_{\mb{C}-\mathrm{field}} = &\mbc{H}_m\otimes\mb{I}_{N_p}-\frac{1}{\varepsilon_0}\mb{d}\otimes\mb{D}\nonumber\\
&\frac{1}{\varepsilon_0}f^2\nu\mb{d}^2\otimes\mb{I}_{N_p}+\nu \mb{I}_{N_m}\otimes \bm{\mathfrak{c}}^{\dagger}\cdot\bm{\mathfrak{c}},\label{eq:HCfieldmatrix}
\end{align}
where $\mb{D}=\curl\mb{C}=-if(\bm{\mathfrak{c}}^{\dagger}-\bm{\mathfrak{c}})$, $\bm{\mathfrak{c}}$ is the $N_{\mathrm{p}}$-dimensional $\mb{C}$-field photon annihilation matrix and the dipole matrix $\mb{d}=-e\mb{x}$ where $\mb{x}=\sum_{n,m=1}^{N_m}x_{n,m}\proj{\epsilon_n}{\epsilon_m}$ with elements
\begin{equation}
x_{n,m}=\begin{cases} -\frac{8L}{\pi^2}\frac{nm}{(n^2-m^2)^2} & n+m\text{ odd} \\ 0 & n+m\text{ even } (n\neq m)\\ L/2 & n=m. \end{cases}
\end{equation}
In Figure~\ref{fig:2} we compare the error in truncating the matter Hilbert spaces to $N_m=2$, which is the 2LT, for the Coulomb and $\mb{C}$-field $(\mb{V},V_0)=(\mb{0},0)$ Hamiltonians. To do so we calculate the energy difference of the two lowest eigenstates in the matrices of Eqns.~\eqref{eq:HCoulombmatrix} and \eqref{eq:HCfieldmatrix}, working in units $\hbar=1=\varepsilon_0$. For both Hamiltonians we do this for $N_m=2$ and $N_m$ large enough for convergence of this energy transition. The latter is the same for both gauges and gives the exact gauge-independent value for this energy transition. For all calculations $N_p$ is also large enough such that the transition is converged with respect to this.

\section{Discussion}\label{sec:discuss}
We stress that we are free to work in any $\mb{C}$-field gauge without affecting the light/matter partitioning, which may offer additional freedom. The $\mb{C}$-field gauge should also be chosen to reflect the physical situation: for instance if the system centre-of-mass is moving a Lorenz gauge is appropriate, whereas a Coulomb gauge is a good choice for static systems. The latter may be useful also if boundaries between different regions are considered, which is not the case for $\mb{A}$-field Coulomb gauge, where a generalisation is required to make the problem tractable \cite{eberlein2019}. In the example of a small dipole in a cavity, the $\mb{C}$-field aligns with the multipolar gauge in the $\mb{A}$-field representation, and so we agree with the conclusion of Refs.~\cite{de2018breakdown,de2018cavity,di2019resolution} that this $\mb{A}$-field gauge choice most accurately represents the physics of small, bound dipoles. 

The equivalence between the $\mb{C}$-field and $\mb{A}$-field approaches warrants further consideration. For example, what choice of $\mb{V}$ and $V_0$ leads to a $\mb{C}$-field Hamiltonian that is analogous to the Coulomb gauge of the $\mb{A}$-field approach? Additionally, it would be interesting to find the set of $(\mb{V},V_0)$ transformations that are allowed, i.e. that cause Eqn.~\eqref{eq:constraint} to vanish.

In conclusion, we find that for systems without free currents and where a truncation of the matter system to few levels is desirable -- typical of cavity QED situations -- the $\mb{C}$-field representation is an attractive choice: it completely removes the dependence of physical predictions after a level truncation on the choice of gauge. In other words, in the $\mb{C}$-field representation, a gauge transformation does not set the light/matter partition. Instead this freedom is moved into the choice of $\mb{P}\pp$ and $\mb{M}$, which may be a more attractive choice. The $\mb{C}$-field approach to QED offers an alternative route, distinct from gauge-fixing in the conventional $\mb{A}$-field representation, to choosing the correct light/matter partition for a given system. In any case, the accuracy of results obtained in the (matter-truncated) $\mb{C}$-representation are independent of the choice of gauge and limited only by the validity of the few-level truncation, decided by the chosen light/matter partition and the system being modelled. 

\begin{acknowledgments}
We thank Adam Stokes, Jonathan Keeling, Christy Kelly, Leone Di Mauro Villari, Ahsan Nazir, Frances Crimin and Stephen M. Barnett for useful discussions. DMR was supported by the UK EPSRC Grant No. EP/L015110/1. EMG acknowledges support from the Royal Society of Edinburgh and Scottish Government and UK EPSRC Grant No. EP/T007214/1. NW wishes to acknowledge financial support from UK EPSRC Grant No. EP/R513222/1 and EP/R030413/1.
\end{acknowledgments}


%

\clearpage

\onecolumngrid
\setcounter{equation}{0}
\renewcommand{\theequation}{SM~\arabic{equation}}
\renewcommand{\thefigure}{SM~\arabic{figure}}    
\renewcommand{\thetable}{SM~\arabic{table}}    
\onecolumngrid
\section*{Supplementary Material  1: The Coulomb and the multipolar gauges for the simple dipole system}\label{app:conv}
We start from the arbitrary-gauge Hamiltonian given in Eqn.~(9) of the main text (repeated here for ease):
\begin{equation}\label{eq:Hminimal}
H\p=\sum_{\mu}\frac{1}{2m_{\mu}}\bigg[\mb{p}\p_{\mu} - q_{\mu}\mb{A}(\mb{r}_{\mu})+q_{\mu}\bN\chi(\mb{r}_{\mu})\bigg]^2+\intx\left(\frac{1}{2\varepsilon_0}\left[\mb{\Pi}\p\x+\bm{\phi}\p\x\right]^2 + \frac{\mb{B}^2\x}{2\mu_0}\right).\nonumber
\end{equation} 

Despite superficial differences the theory is still gauge invariant and so, irrespective of the choice of $\chi$-field, $H_m\p$ will describe the same physics. From Eqn.~(5b) it is clear that the transverse part of the vector potential is gauge invariant. This is quantised in the usual way by satisfying the commutation relation ${\left[\widehat{A}^{\prime\perp}_i\qx,\widehat{\Pi}_j^{\prime\perp}\qxp\right]=i\delta_{ij}^{\perp}(\mb{x}-\mb{x}\p)}$, where $\delta_{ij}^{\perp}(\mb{x}-\mb{x}\p)$ is the transverse $\delta$-function, yielding 
\begin{equation}\label{eq:A}
{\mb{\widehat{A}}}\pp\qx=\sum\kl \bm{\epsilon}\kl f\uk\left({\hat{a}}\kl^{\prime\dagger}\rme^{-i\mb{k}\cdot\mb{x}}+{\hat{a}}\kl\p\rme^{i\mb{k}\cdot\mb{x}}\right),
\end{equation} 
where $f\uk=(2\nu\uk V)^{-{1/2}}$ is the coupling strength to the mode with frequency $\nu\uk=c|\mb{k}|$ in volume $V$, $\bm{\epsilon}\kl$ are polarisation unit vectors orthonormal to $\mb{k}$, with ${\hat{a}}\kl\p$ (${\hat{a}}\kl^{\prime\dagger}$) being a gauge-dependent photon annihilation (creation) operator \cite{bruus2004many,kok2010introduction,mahan2013many,stokes2019gauge}. In the following we will henceforth derive equations for the simple dipole system in the main text instead of the generic charge distribution. From Eqn.~\eqref{eq:A} the magnetic field follows as 
\begin{align}\label{eq:B}
    \widehat{\mb{B}}\qx &=\curl\widehat{\mb{A}}\qx  \\ 
    &=-i\sum\kl \left(\mb{k}\times\bm{\epsilon}\kl\right) f\uk\left({\hat{a}}\kl^{\prime\dagger}\rme^{-i\mb{k}\cdot\mb{x}}-{\hat{a}}\kl\p\rme^{i\mb{k}\cdot\mb{x}}\right).\nonumber
\end{align}
The canonical momentum of the field is gauge dependent, but has the same expression in each gauge found through the commutator ${\left[\widehat{A}^{\prime\perp}_i\qx,\widehat{\Pi}_j^{\prime\perp}\qxp\right]=i\delta_{ij}^{\perp}(\mb{x}-\mb{x}\p)}$, where $\delta_{ij}^{\perp}(\mb{x}-\mb{x}\p)$, and is given by
\begin{equation}\label{eq:Pi}
    \widehat{\bm{\Pi}}^{\prime\perp}\qx=i\sum\kl \nu\uk\bm{\epsilon}\kl f\uk\left({\hat{a}}\kl^{\prime\dagger}\rme^{-i\mb{k}\cdot\mb{x}}-{\hat{a}}\kl\p\rme^{i\mb{k}\cdot\mb{x}}\right).
\end{equation}

\subsection*{A. Coulomb gauge}
\noindent The Coulomb gauge Hamiltonian is defined by choosing $\bN\chi=\mb{A}\pll$ and results in $\bm{\phi}_C=-\varepsilon_0\mb{E}\pll$ and so $\bm{\Pi}_C=-\varepsilon_0\mb{E}\pp$ where we have replaced superscript `prime' with subscript `$C$' to denote the gauge \cite{babiker1983derivation}. The photon field is purely transverse with $\mb{A}_C=\mb{A}\pp$ with conjugate field $\bm{\Pi}_C=-\varepsilon_0\mb{E}\pp$. This also means that the Coulomb gauge canonical momentum is 
\begin{equation}
\mb{p}_{C,\mu}=m_{\mu}\dot{\mb{r}}_{\mu}-q_{\mu}\mb{A}\pp(\mb{r}_{\mu}).\label{eq:pC}
\end{equation}
On substituting these into the arbitrary-gauge Hamiltonian Eqn.~\eqref{eq:Hminimal} for the simple system described in the main text, one finds that
\begin{equation}\label{eq:HmC}
H_{C}=\frac{1}{2m}\bigg[\mb{p}_{C}+e\mb{A}\pp(\mb{r})\bigg]^2+U_{\mathrm{ext}}+U_b +\frac{1}{2}\intx\left(\frac{1}{\varepsilon_0}\left[\mb{\Pi}_C\pp\x\right]^2+\frac{1}{\mu_0}\left[\mb{B}\x\right]^2\right),\nonumber
\end{equation}
where $U_{\mathrm{ext}}$ is an externally applied potential and $U_b=\frac{\varepsilon_0}{2}\intx[\mb{E}\pll\x]^2$ is the electron-hole electrostatic potential which contains the divergent self-energies and interaction between the particles. This potential follows from Gauss's law Eqn.~(1c) and the lack of free charge, which leads to
\begin{align}
\mb{E}\pll\x&=-\bN\intxp\frac{\rho_b\xp}{4\pi\varepsilon_0\left|\mb{x}-\mb{x}\p\right|}\nonumber\\
&\equiv-\frac{1}{\varepsilon_0}\bN V_b\x,
\end{align}
where $V_b=V_e+V_h$ is the sum of electron and hole potentials. Integration by parts, assuming fields vanish at infinities, and again using Gauss's law leads to
\begin{align}
U_b&=\frac{1}{2\varepsilon_0}\intx\rho_b\x V_b\x\nonumber\\
&\equiv U_{e-h}+U_{e-e}+U_{h-h},\label{eq:Veh}
\end{align}
where $U_{e-e}$ and $U_{h-h}$ are the divergent self energies of the electron and hole and $U_{e-h}$ is the electrostatic energy due to the mutual attraction. The $\mb{A}\pp$, $\mb{B}$ and $\bm{\Pi}_C\pp$ fields are quantised as in Eqns.~\eqref{eq:A}, \eqref{eq:B} and \eqref{eq:Pi} albeit with the arbitrary gauge ladder operators replaced with Coulomb-gauge ones. Using these, the terms inside the integral in Eqn.~\eqref{eq:HmC} becomes the usual bosonic field energy term and so the Coulomb-gauge Hamiltonian becomes
\begin{equation}
H_{C}=\frac{1}{2m}\left[\widehat{\mb{p}}_{C}+e\widehat{\mb{A}}\pp(\mb{r})\right]^2 +\widehat{U}_{\mathrm{ext}}+\widehat{U}_b+\sum\kl\nu\uk\left( a\kl^{\dagger}a\kl+\frac{1}{2}\right).
\end{equation} 
In the Coulomb gauge, the light/matter coupling is between the dipole canonical momentum and the gauge invariant transverse vector potential, of the form $\widehat{\mb{p}}_{C}\cdot\widehat{\mb{A}}\pp$. From Eqn.~(7b), we see that the field canonical momentum is proportional to the transverse electric field, $\bm{\Pi}_C=-\varepsilon_0\mb{E}\pp$. We now expand in the basis of $N+1$ energy eigenstates. The system Hamiltonian is formed from 
\begin{equation}
\frac{\widehat{\mb{p}}^{2}_C}{2m}+\widehat{U}_{\mathrm{ext}}\to\sum_{n=0}^N\epsilon_n\left|\epsilon_{n,C}\right>\left<\epsilon_{n,C}\right|\equiv H_{0,C}.\label{eq:pqaunt2}
\end{equation}
Finally, the Coulomb-gauge light matter interaction is written in this basis as follows
\begin{eqnarray}
\frac{e}{m}\widehat{\mb{p}}\cdot\widehat{\mb{A}}\pp&&=ie\left[H_{0,C},\hat{\mb{r}}\right]\cdot\widehat{\mb{A}}\pp\nonumber\\
&&=-\sum_{n=0}^N\sum_{n\p>n}^{N-1}\epsilon_{nn\p}\mb{d}_{nn\p}\cdot\widehat{\mb{A}}\pp\hat{\sigma}^y_{nn\p},\label{eq:VC}
\end{eqnarray}
where $\epsilon_{nn\p}\equiv\epsilon_{n\p}-\epsilon_n$, $\mb{d}_{nn\p}=-e\left<\epsilon_{n,C}\right|\hat{\mb{r}}\left|\epsilon_{n\p,C}\right>$ and $\hat{\sigma}^y_{nn\p}=-i\left(\proj{\epsilon_{n\p,C}}{\epsilon_{n,C}}-\proj{\epsilon_{n,C}}{\epsilon_{n\p,C}}\right)$. In the second line of Eqn.~\eqref{eq:VC} we have written $\mb{p}=m\dot{\mb{r}}$, which is only true in Coulomb gauge in the absence of a vector potential, and is why the unperturbed Hamiltonian is used in the expectation value $\dot{\mb{r}}=i[H_0,\mb{r}]$. Finally, the Coulomb-gauge Hamiltonian written in the energy eigenbasis is
\begin{equation}
H_{C}=\sum_{n=0}^N \epsilon_n\left|\epsilon_{n,C}\right>\left<\epsilon_{n,C}\right|-\sum_{n=0}^N\sum_{n\p>n}^{N-1}\epsilon_{nn\p}\mb{d}_{nn\p}\cdot\widehat{\mb{A}}\pp\hat{\sigma}^y_{nn\p}+\frac{1}{2m}\left[e\widehat{\mb{A}}\pp\right]^2+\sum\kl\nu\uk \hat{a}\kl^{\dagger}\hat{a}\kl,
\end{equation}
where the ladder operators are implicitly in Coulomb gauge. We have made the electric dipole approximation which allows us to evaluate the vector potential at the centre of the dipole at the origin at the coordinate system, i.e. $\mb{A}\pp\equiv\mb{A}\pp(0)$. Additionally, we have ignored the vacuum energy of the photon field, and the divergent self energies of the electron and hole, $\widehat{U}_{e-e}$ and $\widehat{U}_{h-h}$, and also the electron-hole interaction $\widehat{U}_{e-h}$ which is itself divergent in the electric dipole approximation.

\subsection*{B. Multipolar gauge}
\noindent The derivation of the {multipolar} gauge Hamiltonian is more involved and the reader is pointed to \cite{babiker1983derivation,rousseau2017quantum} for the details. It can be summarised by the choice $\bN\chi=\mb{A}\pp-\bm{\Phi}_0/e$. This leads to $\bm{\phi}_m=\mb{P}$ and therefore $\bm{\Pi}_m=-\mb{D}=-\mb{D}\pp$ where the lack of longitudinal component follows from Gauss' law because there are no free charges. Subscript `$m$' labels quantities in the multipolar gauge. By substitution into Eqn.~(9), one finds that the Hamiltonian in the {multipolar} gauge is
\begin{equation}\label{eq:HmM}
H_m=\frac{1}{2m}\left[\mb{p}_{m}+\bm{\Phi}\rr\right]^2+U_{\mathrm{ext}}+\frac{1}{2}\intx\left(\frac{1}{\varepsilon_0}\left[\mb{\Pi}_m\pp\x+\mb{P}\xr\right]^2+\frac{1}{\mu_0}\mb{B}^2\x\right).
\end{equation}
The EM fields are quantised in the same way as in Coulomb-gauge, despite $\bm{\Pi}\pp_m$ and $\bm{\Pi}\pp_C$ corresponding to different physical fields. In the {multipolar} gauge, the electrostatic interaction is governed by the longitudinal component of the polarisation field $\widehat{U}_b=\frac{1}{2\varepsilon_0}\intx[\widehat{\mb{P}}\pll\x]^2$, which is clear from Eqn.~(12a) and leads to the same expression given in Eqn. \eqref{eq:Veh}. In the multipolar gauge, the photon field is much more complicated, $\bm{\Pi}_m=-\mb{D}\pp=-\mb{E}\pp-\mb{P}\pp$, and contains both dipole and transverse field degrees of freedom. The light/matter coupling is between the mechanical dipole position and the canonical momentum of the field, $\intx\widehat{\mb{P}}\pp\x\cdot\widehat{\mb{D}}\pp\x$. Now expanding the matter into the energy basis, the interaction becomes
\begin{align}
\intx\widehat{\mb{P}}\pp\xr \cdot\widehat{\mb{\Pi}}_m\pp\x&=-e\intx\intl\hat{\mb{r}}\cdot\widehat{\mb{\Pi}}_m\pp\x\delta(\mb{x}-\lambda\mb{r})\nonumber\\
&=-e\hat{\mb{r}}\cdot\intl\widehat{\mb{\Pi}}_m\pp(\lambda\mb{r})\nonumber\\
&\approx \widehat{\mb{d}}\cdot\widehat{\mb{\Pi}}_m\pp(0),
\end{align}
where we used the dipole operator $\widehat{\mb{d}}=-e\hat{\mb{r}}$ and in the last line we have made the electric dipole approximation. Within the EDA, the dipolar self-energy term can be rewritten using the transverse delta function as \cite{stokes2019gauge}
\begin{equation}
\frac{1}{2}\intx\left[\widehat{\mb{P}}\pp\xr\right]^2\approx\sum\uk f\uk^2\nu\uk\left(\widehat{\mb{d}}\cdot\bm{\epsilon\kl}\right)^2.
\end{equation}
Applying these, expanding the kinetic energy terms of the eigenbasis as in Eqn. \eqref{eq:pqaunt2} and ignoring the magnetic contribution in $\mb{\Phi}$, we find the standard multipolar Hamiltonian
\begin{equation}\label{eq:HM}
H_{m}=\sum_{n=0}^N \epsilon_n\left|\epsilon_{n,m}\right>\left<\epsilon_{n,m}\right|-\frac{1}{\varepsilon_0}\widehat{\mb{d}}\cdot\widehat{\mb{D}}\pp(0)+\frac{1}{\varepsilon_0}\sum f\uk^2\nu\uk\left(\widehat{\mb{d}}\cdot\bm{\epsilon\kl}\right)^2+\sum\kl\nu\uk \hat{a}\kl^{\dagger}\hat{a}\kl,
\end{equation}
where the ladder operators are implicitly in multipolar gauge. The similarities between the multipolar Hamiltonian derived here in the $\mb{A}$-field approach and the gauge-invariant Hamiltonian we have derived in the new $\mb{C}$-field approach in Eqn.~(26) should be noted (discussed further in the main text). Furthermore, the expansion of the dipole operator into the $N+1$ energy eigenstates is
\begin{equation}
\widehat{\mb{d}}=\sum_{n=0}^N\sum_{n\p>n}^{N-1}\mb{d}_{nn\p}\hat{\sigma}^x_{nn\p},
\end{equation}
where we have assumed that the binding potential gives rise to eigenstates for which the diagonal elements of the dipole transition matrix are zero and $\mb{d}_{nn\p}=-e\left<\epsilon_{n,m}\right|\mb{r}\left|\epsilon_{n\p,m}\right>$. Note that the coupling strength between levels $n$ and $n\p$ scales like $d_{nn\p}\nu$ in the {multipolar} gauge and $d_{nn\p}\epsilon_{nn\p}$ in Coulomb gauge. The increasing coupling strength with energy level separation in Coulomb gauge is related to the breakdown of gauge invariance for finite dipole level truncation, as reported by Refs.~\cite{de2018cavity,de2018breakdown}.

\section*{Supplementary Material 2: Adding an ionic lattice in the new approach}\label{app:ionlattice}
\noindent An ionic lattice can be added to the new approach consistently under the assumption that it does not contribute any macroscopic current. This is justified given that ions vibrate about a mean position. To do so, we note that the partitioning of $\rho$ and $\mb{J}$ into free and bound charges was arbitrary. We could just as well make the partition $\rho=\rho_d+\rho_i$ and $\mb{J}=\mb{J}_d+\mb{J}_i$
where subscripts `$d$' and `$i$' denote charges that are bound in dipoles, and charges that are ions, respectively. The lattice is described by ions of charges $Q_k$ at positions $\mb{R}_k$ with charge density
\begin{equation}
\rho_i\xR=\sum_kQ_k\delta(\mb{x}-\mb{R}_k),
\end{equation}
and $\mb{J}_i=0$. Note that to conserve charge without a current density, the charge density must not be an explicit function of time.
	
A field $\bm{\mathcal{P}}$, analogous to the polarisation field $\mb{P}$, is then defined to be sourced by the dipoles: $\dive\bm{\mathcal{P}}=-\rho_d$ and a field $\bm{\mathcal{D}}$ analogous to the displacement field $\mb{D}$ to be sourced by the ions: $\dive\bm{\mathcal{D}}=\rho_i$. If there are no ions, then the new $\mb{C}$-field theory outlined in the main text is unchanged so long as $\mb{P}\to\bm{\mathcal{P}}$ and $\mb{D}\to\bm{\mathcal{D}}$ because $\dive\mb{E}=\rho$ is still satisfied for $\mb{E}=(\bm{\mathcal{D}}-\bm{\mathcal{P}})/\varepsilon_0$. Including the ions, the Maxwell equations Eqns.~(13) become
\begin{subequations}\label{eq:MLc}
\begin{eqnarray}
&&\dive \bm{\mathcal{D}}\pll\xR=\rho_i\xR,\label{eq:Mc3}\\
&&\curl \mb{H}\x=\dot{\bm{\mathcal{D}}}\pp\x,\label{eq:Mc4}
\end{eqnarray}
\end{subequations}
where we have set $\mb{J}_i=0$. Analogously to the main text, we write the fields in terms of vector potentials to satisfy Eqns. \eqref{eq:MLc}. As before $\mb{H}=\bN C_0+\dot{\mb{C}}$, however, the displacement field analogue requires a longitudinal component
\begin{equation}
\bm{\mathcal{D}}\xR=\curl\mb{C}\x+\bm{\mathcal{D}}\pll\xR,
\end{equation}
with the restrictions that $\dive\bm{\mathcal{D}}\pll=\rho_i$ and $\dot{\bm{\mathcal{D}}}\pll=0$. The latter restriction comes from Eqn.~\eqref{eq:Mc3} and the lack of a macroscopic free current. Note that as before $\bm{\mathcal{D}}\pp\x=\curl\mb{C}\x$. These restrictions are satisfied by choosing
\begin{align}
\bm{\mathcal{D}}\pll\xR&=-\bN\intxp\frac{\rho_i\xpR}{4\pi\left|\mb{x}-\mb{x}\p\right|}\nonumber\\
&\equiv-\bN V_i\xR,\label{eq:Dpll}
\end{align}
where $V_i$ is the electrostatic potential of the ionic lattice. The inclusion of an ionic lattice to the Lagrangian Eqn.~(18) brings an additional kinetic term
\begin{equation}
L\to \bar{L}= L+\sum_k\frac{1}{2}M_k\dot{\mb{R}}^2_k ,
\end{equation}
along with the replacements $\mb{P}\to\bm{\mathcal{P}}$ and $\mb{D}\pp\to\bm{\mathcal{D}}\pp+\bm{\mathcal{D}}\pll$ where $\bm{\mathcal{D}}\pp=\mb{D}\pp=\curl\mb{C}$. In order for the lattice to have been added consistently we must be able to derive the Lorentz force acting on the ions by using the Euler-Lagrange equation with ion position $\mb{R}_k$. The velocity derivative clearly results in $\partial \bar{L}/\partial\dot{\mb{R}}_k=M_k\dot{\mb{R}}_k$, though the position derivative is more difficult. To start we rewrite Eqn. \eqref{eq:Dpll} as
\begin{align}
	\bm{\mathcal{D}}\pll\xR &=-\sum_kQ_k\nabla\frac{1}{4\pi\left|\mb{x}-\mb{R}_k\right|}\nonumber\\
	&=\sum_kQ_k\Dfrac{}{\mb{R}_k}\frac{1}{4\pi\left|\mb{x}-\mb{R}_k\right|}.
\end{align}
We can then write the ion position derivative of the Lagrangian as
\begin{align}
\Dfrac{\bar{L}}{\mb{R}_k}&=\Dfrac{}{\mb{R}_k}\intx\left(\frac{\bm{\mathcal{D}}^2\xR}{2}+\bm{\mathcal{D}}\xR\cdot\bm{\mathcal{P}}\xr\right)\nonumber\\
&=-\intx\left(\mb{E}\x\cdot\Dfrac{}{\mb{R}_k}\right)\bm{\mathcal{D}}\pll\xR,
\end{align}
where we have enforced that $\bm{\mathcal{P}}$ and $\bm{\mathcal{D}}\pp$ are not functions of $\mb{R}_k$ and that $\partial/\partial\mb{R}_k\times\bm{\mathcal{D}}\pll=0$ because $\bm{\mathcal{D}}\pll$ is a longitudinal field with respect to $\mb{R}_k$. The $j$-th component of the integrand is therefore 
\begin{equation}
\bigg[\left(\mb{E}\x\cdot\Dfrac{}{\mb{R}_k}\right)\bm{\mathcal{D}}\pll\xR\bigg]_j=\;Q_k
\left(E_i\x\Dfrac{}{R_{k,i}}\right)\Dfrac{}{R_{k,j}}\left(\frac{1}{4\pi\left|\mb{x}-\mb{R}_k\right|}\right),
\end{equation}
where the sum over $i$ is implied. The longitudinal and transverse delta functions are \cite{stewart2008longitudinal}
\begin{align}
\delta\pll_{ij}\qx &=-\lim_{\epsilon\to0}\Dfrac{}{x_i}\Dfrac{}{x_j}\frac{1}{4\pi\sqrt{\mb{x}^2+\epsilon^2}}\nonumber\\
&=\frac{1}{3}\delta_{ij}\qx+\frac{1}{4\pi\left|\mb{x}\right|^3}\left(\delta_{ij}-3\frac{x_ix_j}{\left|\mb{x}\right|^2}\right)\ 
\end{align}
and
\begin{align}
\delta_{ij}\pp\qx=\delta_{ij}\qx-\delta\pll_{ij}\qx,
\end{align}
where $\delta_{ij}$ is a Kronecker-delta and $\delta_{ij}\qx\equiv\delta_{ij}\delta\qx$. Therefore,
\begin{align}
\left[\left(\mb{E}\x\cdot\Dfrac{}{\mb{R}_k}\right)\bm{\mathcal{D}}\pll\xR\right]_j=-Q_kE_i(x)\delta_{ij}\pll\left(\mb{x}-\mb{R}_k\right),
\end{align}
which gives the Lorentz force
\begin{equation}
M_k\ddot{\mb{R}}_k=Q_k \mb{E}\pll(\mb{R}_k).
\end{equation}
The lack of magnetic force and transverse electric force originates from assuming that the ions do not generate any current. Therefore, the ions are not affected by magnetic forces and the electrostatic force is conservative which means it has zero curl and so no transverse component. 
	
Since $\dot{\bm{\mathcal{D}}}\pll=0$, the mechanical degree of freedom $\bm{\mathcal{D}}\pll$ does not have a canonical momentum. The ion position does, however, have a canonical momentum given by
\begin{equation}
\mb{P}_k=\Dfrac{\bar{L}}{\dot{\mb{R}}_k}=M_k\dot{\mb{R}}_k,
\end{equation}
which brings an additional kinetic energy term to the total Hamiltonian $\bar{H}$. This Hamiltonian is then found to be (using $\mb{V}=V_0=0$)
\begin{align}\label{eq:HnewPhonon}
\bar{H}=&\frac{1}{2m}\left[\mb{p}+\bm{\Phi}(\mb{r})\right]^2+\sum_k\frac{\mb{P}_k^2}{2M_k} +\frac{1}{2}\intx\left(\frac{1}{\varepsilon_0}\left[\bm{\mathcal{D}}\pp\x-\bm{\mathcal{P}}\xr\right]^2+\frac{1}{\mu_0}\mb{B}^2\x\right)\nonumber\\ &+\frac{1}{\varepsilon_0}\intx\bigg(\frac{1}{2}\left[\bm{\mathcal{D}}\pll\xR\right]^2+\bm{\mathcal{D}}\pll\xR\cdot\bm{\mathcal{P}}\xr\bigg).
\end{align}
As described in the main text, we can quantise the $\mb{C}\pp$-field as in Eqn.~(25). The radiation fields are quantised exactly as without the ionic lattice. That is, for the $\mb{C}\pp$-field quantised as in Eqn.~(25), $\bm{\mathcal{D}}\pp$ takes the form
\begin{equation}
    \widehat{\bm{\mathcal{D}}}\pp\qx=-i\sum\kl \left(\mb{k}\times\bm{\epsilon}\kl\right) f\uk\big({\hat{\mathfrak{c}}}\kl^{\dagger}\rme^{-i\mb{k}\cdot\mb{x}}
    \hspace{0.2cm}-{\hat{\mathfrak{c}}}\kl\rme^{i\mb{k}\cdot\mb{x}}\big),
\end{equation}
and the magnetic field is just the canonical momentum. Therefore, using the canonical commutation relation $\left[\widehat{C}_i\qx,\widehat{B}_j\qxp\right]=\left[\widehat{C}_i\qx,\widehat{B}_j\pp\qxp\right]=i\delta^\perp_{ij}(\mb{x}-\mb{x}\p)$ where the second equality follows from $\dive\mb{B}=0$, we find that
\begin{equation}
    \widehat{\mb{B}}\x=i\sum\kl \nu\uk\bm{\epsilon}\kl f\uk\left({\hat{\mathfrak{c}}}\kl^{\dagger}\rme^{-i\mb{k}\cdot\mb{x}}-{\hat{\mathfrak{c}}}\kl\rme^{i\mb{k}\cdot\mb{x}}\right).
\end{equation}
After quantising the radiation fields in this way, it is clear that the first line of Eqn.~\eqref{eq:HnewPhonon} will lead to the Hamiltonian describing the dipole interacting with the light field that is given in the main text in Eqn.~(26), so long as the magnetic effects in $\bm{\Phi}$ are ignored and the electric dipole approximation is made to simplify $\bm{\mathcal{P}}\x\approx-e\mb{r}\delta(\mb{x})$. Quantising the second line of Eqn. \eqref{eq:HnewPhonon} leads to a description of the influence (on the dipole) of vibrations in the ionic lattice from their mean positions. From Eqn.~\eqref{eq:Dpll} we can see that the total energy of the phonon field is described by the terms
\begin{equation}
\sum_k\frac{\mb{P}_k^2}{2M_k}+\intx\frac{1}{2\varepsilon_0}\left[\widehat{\bm{\mathcal{D}}}\pll\xR\right]^2=\sum_{\mb{q}\mu}\omega_{\mb{q}\mu}\left(\hat{b}^{\dagger}_{\mb{q}\mu}\hat{b}_{\mb{q}\mu}+\frac{1}{2}\right),\label{eq:phononE}
\end{equation}
where $b_{\mb{q}\mu}$ ($b^{\dagger}_{\mb{q}\mu}$) is the annihilation (creation) operator of a phonon of wavenumber $\mb{q}$ and polarisation $\mu$. Details of the derivation of Eqn. \eqref{eq:phononE} can be found in, for example, Refs. \cite{kok2010introduction,mahan2013many,bruus2004many,nazir2016modelling}. The final term in the Hamiltonian, $\intx\widehat{\bm{\mathcal{D}}}\pll\cdot\widehat{\bm{\mathcal{P}}}$, describes the interaction between the phonons and the dipole. After linearising this interaction by assuming that the ions are only displaced from equilibrium by small amounts, one can then show that in the EDA \cite{kok2010introduction,mahan2013many,bruus2004many,nazir2016modelling}
\begin{equation}
\frac{1}{\varepsilon_0}\intx \widehat{\bm{\mathcal{D}}}\pll\xR\cdot\widehat{\bm{\mathcal{P}}}\xr
\approx\sum_{\mb{q}\mu}\left(\hat{b}^{\dagger}_{\mb{q}\mu}+\hat{b}_{\mb{q}\mu}\right)\sum_{n=1}^Ng_{n\mb{q}\mu}\left|\epsilon_{n}\right>\left<\epsilon_{n}\right|,
\end{equation}
which is the usual phonon interaction encountered throughout the literature where the coupling strength of dipole level $n$ to phonon mode with wavenumber $\mb{q}$ and polarisation $\mu$ is $g_{n\mb{q}\mu}$. We can finally write the full Hamiltonian describing the dynamics of a dipole in an EM field interacting with a vibrating ionic lattice (with $\mb{V}=V_0=0$):
\begin{align}
\widehat{H}=&\sum_{n=0}^N\epsilon_n\left|\epsilon_{n}\right>\left<\epsilon_{n}\right|-\frac{1}{\varepsilon_0}\widehat{\mb{d}}\cdot\left(\curl\widehat{\mb{C}}\pp(0)\right)+\frac{1}{\varepsilon_0}\sum f\uk^2\nu\uk\left(\widehat{\mb{d}}\cdot\bm{\epsilon\kl}\right)^2+\sum\kl\nu\uk\nonumber \hat{\mathfrak{c}}\kl^{\dagger}\hat{\mathfrak{c}}\kl\\
&+\sum_{\mb{q}\mu}\left(\hat{b}^{\dagger}_{\mb{q}\mu}+\hat{b}_{\mb{q}\mu}\right)\sum_{n=1}^Ng_{n\mb{q}\mu}\left|\epsilon_{n}\right>\left<\epsilon_{n}\right|+\sum_{\mb{q}\mu}\omega_{\mb{q}\mu}\hat{b}^{\dagger}_{\mb{q}\mu}\hat{b}_{\mb{q}\mu}.
\end{align}

\section*{Supplementary Material 3: Deriving the equations of motion and Hamiltonian from the $\mb{V}=V_0=0$ theory}\label{app:deriv}
In this section, we will derive the equations of motion and the Hamiltonian of the $\mb{C}$-field theory from the Lagrangian given in the main text by Eqn.~(18). For ease of reading, this Lagrangian is
\begin{equation}
L=\frac{1}{2}m\dot{\mb{r}}^2+\intx\frac{1}{2}\bigg(\mu_0\left[\mb{H}\x+\mb{M}\xr\right]^2-\frac{1}{\varepsilon_0}\left[\mb{D}\x-\mb{P}\xr\right]^2 \bigg),\nonumber
\end{equation}
where $\mb{D}=\curl\mb{C}$ and $\mb{H}=\boldsymbol{\nabla}C_0+\dot{\mb{C}}$, which automatically satisfy Gauss's law $\dive\mb{D}=0$ and Maxwell-Ampere's law $\curl\mb{H}=\dot{\mb{D}}$. Note that the equations of motion for the Lagrangian transformed by Eqns.~(17) must be identical because the transformation preserves the equations of motion by construction.
\subsection*{A. No magnetic monopoles}
The non-existence of magnetic monopoles means that $\mb{B}$ is sourceless and hence divergenceless: $\dive\mb{B}=0$. This equation of motion is found from the Euler-Lagrange equation
\begin{equation}
\frac{d}{dt}\Dfrac{L}{\dot{C}_0\x}-\Dfrac{L}{C_0\x}=0.
\end{equation}
Since $L$ does not depend on $\dot{C}_0$ we can immediately write that $\partial L/\partial \dot{C}_0=0$. We can also show that
\begin{align}
	\Dfrac{L}{C_0\x}&=\Dfrac{}{C_0\x}\intxp\mu_0\bigg[\frac{1}{2}\mb{H}^2\xp+\mb{H}\xp\cdot\mb{M}\xpr\bigg]\nonumber\\
	&=\intxp\mu_0\left[\mb{H}\xp+\mb{M}\xpr\right]\cdot\Dfrac{}{C_0\x}\nabla C_0\xp\nonumber\\
	&=-\intxp\left[\dive\mb{B}\xp\right]\Dfrac{C_0\xp}{C_0\x}\nonumber\\
	&=-\dive\mb{B}\x,
\end{align}
where, to arrive at the third line, we have integrated by parts and, to arrive at the final line, we have used that $\partial C_0\xp/\partial C_0\x=\delta(\mb{x}-\mb{x}\p)$. This completes the proof. Note that treating $C_0$ as an independent coordinate results in a singular Hessian and so a non-invertible Lagrangian. It is better to not treat $C_0$ as a coordinate, and instead impose $\dive\mb{B}=0$ as a primary constraint on the Lagrangian. This is identical to Gauss's law $\dive\mb{E}=\rho$ and $A_0$ in the conventional theory.

\subsection*{B. Faraday's equation}
Faraday's equation, $\curl\mb{E}=-\dot{\mb{B}}$, is found from the Euler-Lagrange equation
\begin{equation}
\frac{d}{dt}\Dfrac{L}{\dot{\mb{C}}\x}-\Dfrac{L}{\mb{C}\x}=0.
\end{equation}
We start with the derivative with respect to $\mb{C}$, for which we find that
\begin{align}
	\Dfrac{L}{\dot{\mb{C}}\x}&=\Dfrac{}{\dot{\mb{C}}\x}\intxp\mu_0\left[\frac{1}{2}\mb{H}^2\xp+\mb{H}\xp\cdot\mb{M}\xpr\right]\nonumber\\
	&=\intxp\mb{B}\xp\cdot\Dfrac{\dot{\mb{C}}\xp}{\dot{\mb{C}}\x}\nonumber\\
	&=\mb{B}\x,\label{eq:dLdCdot}
\end{align}
and note that this is also the conjugate momentum of the field. The other derivative is
\begin{equation}
\Dfrac{L}{\mb{C}\x}=\Dfrac{}{\mb{C}\x}\intxp\frac{1}{\varepsilon_0}\bigg[-\frac{1}{2}\mb{D}^2\xp+\mb{D}\xp\cdot\mb{P}\xpr\bigg].
\end{equation}
This is slightly more complicated so we will take each term individually. The $\mb{D}^2$ term is evaluated using $\mb{D}=\curl\mb{C}$ and the identity $\partial/\partial\mb{W}\left(\curl\mb{W}\right)^2=2\curl\curl\mb{W}$ for any vector field $\mb{W}$. The $\mb{D}\cdot\mb{P}$ term is evaluated as
\begin{align}\label{eq:DcdotP}
	&\Dfrac{}{\mb{C}\x}\intxp\mb{D}\xp\cdot\mb{P}\xpr\\
	&\hspace{1cm}=\intxp\bigg[\left(\mb{P}\xpr\cdot\Dfrac{}{\mb{C}\x}\right)\mb{D}\xp+\mb{P}\xpr\times\left(\Dfrac{}{\mb{C}\xp}\times\mb{D}\xp\right)\bigg].\nonumber
\end{align}
For the first term of Eqn. \eqref{eq:DcdotP} we find that
\begin{equation}
\left(\mb{P}\xpr\cdot\Dfrac{}{\mb{C}\x}\right)\mb{D}\xp=\curl\left[\mb{P}\xpr\delta(\mb{x}-\mb{x}\p)\right].
\end{equation}
The second term in Eqn. \eqref{eq:DcdotP} is zero because $\partial/\partial\mb{C}\times\left[\curl\mb{C}\right]=\curl\left[\partial/\partial\mb{C}\times\mb{C}\right]=0$. Therefore, we find that
\begin{align}
	\Dfrac{L}{\mb{C}\x}&=-\curl\intxp\frac{1}{\varepsilon_0}\left[\mb{D}\xp-\mb{P}\xpr\right]\delta(\mb{x}-\mb{x}\p)\nonumber\\
	&=-\curl\mb{E}\x.\label{eq:dLdC}
\end{align}
Together, Eqns. \eqref{eq:dLdCdot} and \eqref{eq:dLdC} complete the proof.

\subsection*{C. Lorentz force}
The proceeding derivation of the Lorentz force is very lengthy. It can be bypassed by noting that our Hamiltonian in Eqn.~(23) is identical to the multipolar Hamiltonian in \cite{babiker1983derivation} when written in terms of the physical fields. Since the multipolar gauge Hamiltonian is equivalent to the Coulomb gauge Hamiltonian, and it is readily proven that the Coulomb gauge Hamiltonian reproduces the Lorentz force through Heisenberg's equation, so must the $\mb{C}$-field Hamiltonian. This argument can only be made for the Lorentz force derivation because this does not depend on the choice of potentials to characterise the physical fields. Nevertheless, for concreteness here we will derive the Lorentz force explicitly from the $\mb{C}$-field Lagrangian.

The Lorentz force of the electron, $m\ddot{\mb{r}}=-e\left[\mb{E}(\mb{r})+\dot{\mb{r}}\times\mb{B}(\mb{r})\right]$, is found from the Euler-Lagrange equation
\begin{equation}
\frac{d}{dt}\Dfrac{L}{\dot{\mb{r}}}-\Dfrac{L}{\mb{r}}=0.\label{eq:ELr}
\end{equation}
Note that to derive the Lorentz force for the hole one must redefine the origin of the coordinate system such that the hole is not located on it. Then, subsequently evaluate the Euler-Lagrange with respect to the new position vector of the hole. To evaluate Eqn.~\eqref{eq:ELr} consider the velocity derivative,
\begin{equation}
\Dfrac{L}{\dot{\mb{r}}}=m\dot{\mb{r}}+\Dfrac{}{\dot{\mb{r}}}\intx\mu_0\bigg[\mb{H}\x \cdot\mb{M}\xr+\frac{1}{2}\mb{M}^2\xr\bigg].
\end{equation}
It can be shown that
\begin{eqnarray}
&&\Dfrac{}{\dot{\mb{r}}}\mb{H}\x\cdot\mb{M}\xr=-\bm{\theta}\xr\times\mb{H}\x,\\ &&\Dfrac{}{\dot{\mb{r}}}\frac{1}{2}\mb{M}^2\xr=-\bm{\theta}\xr\times\mb{M}\xr.
\end{eqnarray}
Therefore,
\begin{eqnarray}
\Dfrac{L}{\dot{\mb{r}}}&&=m\dot{\mb{r}}-\intx\bm{\theta}(\mb{x},\mb{r})\times\mb{B}\x\nonumber\\
&&=m\dot{\mb{r}}+e\intl\lambda\mb{r}\times\mb{B}(\lambda\mb{r}),\label{eq:Lrdot}
\end{eqnarray}
which is also equal to the conjugate momentum $\mb{p}$ in Eqn.~(21a) if $\mb{V}=V_0=0$. We then take the total time derivative of this to find
\begin{align}
	\ddfrac{}{t}\Dfrac{L}{\dot{\mb{r}}}=m\ddot{\mb{r}}+e\intl\lambda\bigg(\dot{\mb{r}}\times\mb{B}+\mb{r}\times\dot{\mb{B}}\lr+\mb{r}\times\left[\left(\dot{\mb{r}}\cdot\nabla_{\mb{r}}\right)\mb{B}\lr\right]\bigg).\label{eq:LorentzLHS}
\end{align}
The position derivative separates into electric and magnetic parts
\begin{equation}
\Dfrac{L}{\mb{r}}=\intx\nabla_{\mb{r}}\left(\mathcal{L}_{\mathrm{elec}}+\mathcal{L}_{\mathrm{mag}}\right),
\end{equation}
where
\begin{align}
	\nabla_{\mb{r}}\mathcal{L}_{\mathrm{elec}}&=\varepsilon_0^{-1}\nabla_{\mb{r}}\left(\mb{D}\x\cdot\mb{P}\xr-\frac{1}{2}\mb{P}^2\xr\right)\nonumber\\
	&=\left[\mb{E}\x\cdot\nabla_{\mb{r}}\right]\mb{P}\xr+\mb{E}\x\times\left[\nabla_{\mb{r}}\times\mb{P}\xr\right]\nonumber\\
	&=\tilde{\nabla}_{\mb{P}}\left[\mb{E}\x\cdot\mb{P}\xr\right],
\end{align}
and the terms involving partial derivatives of $\mb{D}$ with respect to $\mb{r}$ are zero. To arrive at the last line we have used the identity $\mb{a}\times\left(\nabla_{\mb{r}}\times\mb{b}\right)=\tilde{\nabla}_{\mb{b}}\left(\mb{a}\cdot\mb{b}\right)-\left(\mb{a}\cdot\nabla_{\mb{r}}\right)\mb{b}$ where the gradient operator with a tilde is the Feynman subscript notation, i.e.
\begin{equation}
\tilde{\nabla}_{\mb{b}}\left(\mb{a}\cdot\mb{b}\right)=\left(\mb{a}\cdot\Dfrac{\mb{b}}{r_x},\mb{a}\cdot\Dfrac{\mb{b}}{r_y},\mb{a}\cdot\Dfrac{\mb{b}}{r_z}\right).
\end{equation}
Likewise the magnetic part is
\begin{align}
	\nabla_{\mb{r}}\mathcal{L}_{\mathrm{mag}}&=\nabla_{\mb{r}}\left(\mb{H}\x\cdot\mb{M}\xr+\frac{1}{2}\mb{M}^2\xr\right)\nonumber\\
	&=\left[\mb{B}\x\cdot\nabla_{\mb{r}}\right]\mb{M}\xr+\mb{B}\x\times\left[\nabla_{\mb{r}}\times\mb{M}\xr\right]\nonumber\\
	&=\tilde{\nabla}_{\mb{M}}\left[\mb{B}\x\cdot\mb{M}\xr\right],
\end{align}
and partial derivatives of $\mb{H}$ with respect to $\mb{r}$ are zero. The electric part evaluates to
\begin{align}\label{eq:L1}
	\intx\nabla_{\mb{r}}\mathcal{L}_{\mathrm{elec}}&=\intx\tilde{\nabla}_{\mb{P}}\left[\mb{E}\x\cdot\mb{P}\xr\right]\nonumber\\
	&=-e\intl\intx\bigg[\mb{E}\x+\left(\mb{E}\x\cdot\mb{r}\right)\nabla_{\mb{r}}\bigg]\delta\left(\mb{x}-\lambda\mb{r}\right).
\end{align}
and the magnetic part is 
\begin{align}\label{eq:L2}
	\intx\nabla_{\mb{r}}\mathcal{L}_{\mathrm{mag}}&=\intx\tilde{\nabla}_{\mb{M}}\left[\mb{B}\x\cdot\mb{M}\xr\right]\nonumber\\
	&=-e\intl\intx\bigg\{\dot{\mb{r}}\times\mb{B}\x+\left(\mb{r}\cdot\left[\dot{\mb{r}}\times\mb{B}\x\right]\right)\nabla_{\mb{r}}\bigg\}\delta\left(\mb{x}-\lambda\mb{r}\right).
\end{align}
Before substituting into the Euler-Lagrange equation we note that by using Faraday's law we can rewrite the second term in Eqn. \eqref{eq:LorentzLHS} as 
\begin{align}
	e\intl\lambda\mb{r}\times\dot{\mb{B}}\lr&=-e\intl\mb{r}\times\left[\nabla_{\mb{r}}\times\mb{E}\lr\right]\nonumber\\
	&=-e\intl\bigg\{\tilde{\nabla}_{\mb{E}}\left[\mb{r}\cdot\mb{E}\lr\right]-\left[\mb{r}\cdot\nabla_{\mb{r}}\right]\mb{E}\lr\bigg\}\nonumber\\
	&=-e\intl\intx\bigg\{\left[\mb{r}\cdot\mb{E}\x\right]\nabla_{\mb{r}}-\mb{E}\x\left[\mb{r}\cdot\nabla_{\mb{r}}\right]\bigg\}\delta(\mb{x}-\lambda\mb{r}),\label{eq:L3}
\end{align}
where to arrive at the last line we have written $\mb{E}\lr=\intx\mb{E}\x\delta(\mb{x}-\lambda\mb{r})$. Substituting Eqns. \eqref{eq:LorentzLHS}, \eqref{eq:L1}, \eqref{eq:L2} and \eqref{eq:L3} into the Euler-Lagrange equation results in
\begin{align}\label{eq:LorentzHalfWay}
	m\ddot{\mb{r}}=&-e\intl\intx\bigg\{\mb{E}\x\left[1+\left(\mb{r}\cdot\nabla_{\mb{r}}\right)\right]\nonumber\\
	&\qquad+\lambda\bigg[\left[\mb{r}\times\mb{B}\x\right]\left(\dot{\mb{r}}\cdot\nabla_{\mb{r}}\right)+2\dot{\mb{r}}\times\mb{B}+\left(\mb{r}\cdot\left[\dot{\mb{r}}\times\mb{B}\x\right]\right)\nabla_{\mb{r}}\bigg]\bigg\}\delta(\mb{x}-\lambda\mb{r}),
\end{align}
where we have also written $\mb{B}\lr=\intx\mb{B}\x\delta(\mb{x}-\lambda\mb{r})$. After some algebra we can then prove that
\begin{align}
	\left[\mb{r}\times\mb{B}\x\right]\left(\dot{\mb{r}}\cdot\nabla_{\mb{r}}\right)+\left(\mb{r}\cdot\left[\dot{\mb{r}}\times\mb{B}\x\right]\right)\nabla_{\mb{r}}=\left[\dot{\mb{r}}\times\mb{B}\x\right]\left(\mb{r}\cdot\nabla_{\mb{r}}\right)+\left(\mb{r}\times\dot{\mb{r}}\right)\left[\mb{B}\x\cdot\nabla_{\mb{r}}\right],
\end{align}
which when substituted into Eqn. \eqref{eq:LorentzHalfWay} along with $\nabla_{\mb{r}}\delta(\mb{x}-\lambda\mb{r})=-\lambda\nabla_{\mb{x}}\delta(\mb{x}-\lambda\mb{r})$ we find
\begin{align}
	m\ddot{\mb{r}}&=-e\intx\mb{E}\x\intl\left[1-\lambda\left(\mb{r}\cdot\nabla_{\mb{x}}\right)\right]\delta(\mb{x}-\lambda\mb{r})\nonumber\\
	&-e\intx\ \left[\dot{\mb{r}}\times\mb{B}\x\right]\intl\lambda\left(2-\lambda\mb{r}\cdot\nabla_{\mb{x}}\right)\delta(\mb{x}-\lambda\mb{r})\nonumber\\
	&-\intl\lambda^2\left(\mb{r}\times\dot{\mb{r}}\right)\intx\left[\mb{B}\x\cdot\nabla_{\mb{x}}\right]\delta(\mb{x}-\lambda\mb{r}).\nonumber
\end{align}
From here the Lorentz force is derived by noting that because of the non-existence of magnetic monopoles
\begin{equation}
\intx\left[\mb{B}\x\cdot\nabla_{\mb{x}}\right]\delta(\mb{x}-\lambda\mb{r})=-\intx\left[\nabla_{\mb{x}}\cdot\mb{B}\x\right]\delta(\mb{x}-\lambda\mb{r})=0,
\end{equation}
and that the following identities hold
\begin{align}
	\intl\left(1-\lambda\mb{r}\cdot\nabla_{\mb{x}}\right)\delta(\mb{x}-\lambda\mb{r})&=\delta(\mb{x}-\mb{r}),\\
	\intl\lambda\left(2-\lambda\mb{r}\cdot\nabla_{\mb{x}}\right)\delta(\mb{x}-\lambda\mb{r})&=\delta(\mb{x}-\mb{r}).
\end{align}
These identities can be proven using the Fourier representation of the delta function \cite{babiker1983derivation}. Using these we find that $m\ddot{\mb{r}}=-e\left[\mb{E}(\mb{r})+\dot{\mb{r}}\times\mb{B}(\mb{r})\right]$ as required.

\subsection*{D. Hamiltonian}
The Hamiltonian is derived as is standard by
\begin{equation}\label{eq:Hgen}
H=\sum_i\mb{p}_i\cdot\dot{\mb{q}}_i-L,
\end{equation}
where $\mb{q}_i$ and $\mb{p}_i$ are the generalised coordinates and canonical momenta. For this theory, the canonical momenta of the matter and $\mb{C}$-field are given in Eqns.~(21) with $\mb{V}=V_0=0$. We find that
\begin{equation}
H=\mb{p}\cdot\dot{\mb{r}}-\frac{1}{2}m\dot{\mb{r}}^2+\intx\left(\mb{B}\x\cdot\dot{\mb{C}}\x-\frac{\mu_0}{2}\mb{B}^2\x+\frac{\varepsilon_0}{2}\mb{E}^2\x\right).
\end{equation}
Focusing first on the terms within the integral, we can rewrite $\mb{B}\cdot\dot{\mb{C}}=\mb{B}\cdot\mb{H}-\mb{B}\cdot\nabla C_0$. The term $\mb{B}\cdot\nabla C_0$ then vanishes after integration by parts and enforcing $\dive\mb{B}=0$. Since the canonical field momentum is $\mb{B}$, we want to eliminate $\mb{H}$ in favour of $\mb{B}$ to make the quantisation process easier. Therefore, writing $\mb{H}+\mb{M}=\mb{B}/\mu_0$ we arrive at
\begin{equation}
H=\mb{p}\cdot\dot{\mb{r}}-\frac{1}{2}m\dot{\mb{r}}^2+\frac{1}{2}\intx\left(\varepsilon_0\mb{E}^2\x+\frac{1}{\mu_0}\mb{B}^2\x-2\mb{B}\x\cdot\mb{M}\xr\right).
\end{equation}
Using the definitions of $\bm{\theta}$ and $\bm{\Phi}_0$ from the main text we are then able to show that
\begin{equation}
\intx\mb{B}\x\cdot\mb{M}\xr=-\frac{1}{m}\left[\mb{p}\cdot\bm{\Phi}_0\rr+\bm{\Phi}_0^2\rr\right].
\end{equation}
Finally, after combining this with the terms outside the integral using $m\dot{\mb{r}}=\mb{p}+\bm{\Phi}_0$, we have
\begin{equation}
H=\frac{1}{2m}\left[\mb{p}+\bm{\Phi}_0\rr\right]^2+\intx \left(\frac{\mb{B}^2\x}{2\mu_0}+\frac{1}{2\varepsilon_0}\left[\mb{D}\pp\x-\mb{P}\xr\right]^2\right),
\end{equation}
as in Eqn.~(23) of the main text. Note that here we have explicitly used that the longitudinal component of $\mb{D}$ is zero because there is no free charge. To align with the Hamiltonians in Supplementary Material~1 for the conventional theory we will ignore the magnetic interactions governed by $\bm{\Phi}_0$. Quantising the radiation fields is described in Supplementary Material~2 and results in Eqn.~(26) in the main text.

\section*{Supplementary Material 4: Deriving the canonical momenta and Hamiltonian with nonzero $\mb{V}$ and $V_0$}\label{app:deriv-V}
\subsection*{A. Field and matter canonical momenta}
After transformation by Eqns.~(17) the Lagrangian is
\begin{equation}
\wt{L}=\frac{1}{2}m\dot{\mb{r}}^2+\intx\wt{\mathcal{L}}\x,
\end{equation}
where
\begin{align}
	\wt{\mathcal{L}}\x&=\frac{1}{2}\left(\frac{1}{\mu_0}\mb{B}^2\x-\varepsilon_0\mb{E}^2\x\right)\\
	&=\frac{1}{2}\bigg(\mu_0\left[\wt{\mb{H}}\x+\wt{\mb{M}}\xr\right]^2-\frac{1}{\varepsilon_0}\left[\wt{\mb{D}}\x-\wt{\mb{P}}\xr\right]^2\bigg).
\end{align}
Also, recall the definitions $\wt{\mb{P}}=\mb{P}+\wt{\mb{P}}_V$, $\wt{\mb{M}}=\mb{M}-\wt{\mb{M}}_V$,
\begin{align}
	&\wt{\mb{P}}_V\xr=\curl\mb{V}\xr,\\
	&\wt{\mb{M}}_V\xr=\dot{\mb{V}}\xr+\bN V_0\xr,\\
	&\wt{\mb{D}}\x=\curl\wt{\mb{C}}\x,\\
	&\wt{\mb{H}}\x=\dot{\wt{\mb{C}}}\x+\bN \wt{C}_0\x,
\end{align}
where we have written the implicit change in $\mb{C}$ ($C_0$) due to the change in light/matter partition explicitly as $\mb{C}\to\wt{\mb{C}}$ ($C_0\to\wt{C}_0$). It is clear that the field canonical momentum will not change from the $\mb{V}=V_0=0$ case, because it is equal to the physical field $\mb{B}$. Therefore we trivially find that
\begin{equation}
\wt{\bm{\Pi}}\x=\Pfrac{\wt{\mathcal{L}}}{\dot{\wt{\mb{C}}}\x}=\mu_0\left[\wt{\mb{H}}\x+\wt{\mb{M}}\xr\right]=\mb{B}\x.
\end{equation}
For the matter canonical momentum we must evaluate
\begin{equation}
\wt{\mb{p}}=m\dot{\mb{r}}+\mu_0\bN_{\dot{\mb{r}}}\intx\left(\wt{\mb{H}}\x\cdot\wt{\mb{M}}\xr+\frac{1}{2}\wt{\mb{M}}^2\xr\right).
\end{equation}
To do so we recall some algebraic relations used in the $\mb{V}=V_0=0$ case:
\begin{subequations}\label{eq:V0IdM}
	\begin{align}
		&\bN_{\dot{\mb{r}}}\left[\mb{W}\x\cdot\mb{M}\xr\right]=-\bm{\theta}\xr\cdot\mb{W}\x\ \forall\ \mb{W}\x,\\
		&\bN_{\dot{\mb{r}}}\frac{1}{2}\mb{M}^2\xr=-\bm{\theta}\xr\cdot\mb{M}\xr.
	\end{align}
\end{subequations}
We can expand the integral as
\begin{equation}
\wt{\mb{p}}=m\dot{\mb{r}}+\mu_0\bN_{\dot{\mb{r}}}\intx\bigg\{\frac{1}{2}\mb{M}^2\x+\frac{1}{2}\wt{\mb{M}}_V^2\xr-\wt{\mb{H}}\x\cdot\mb{M}\xr-\left[\mb{M}\xr+\wt{\mb{H}}\x\right]\cdot\wt{\mb{M}}_V\xr\bigg\}.
\end{equation}
Using Eqns.~\eqref{eq:V0IdM} we can show that
\begin{equation}
\mu_0\bN_{\dot{\mb{r}}}\intx\left(\frac{1}{2}\mb{M}^2\xr-\wt{\mb{H}}\x\cdot\mb{M}\xr\right)=-\mu_0\intx\bm{\theta}\rr\times\left[\mb{M}\xr+\wt{\mb{H}}\x\right].
\end{equation}
Then, using standard vector calculus identities we find that the remaining terms can be written as
\begin{align}
	\mu_0\bN_{\dot{\mb{r}}}\intx\bigg(\frac{1}{2}\wt{\mb{M}}_V^2\xr&-\left[\mb{M}\xr+\wt{\mb{H}}\x\right]\cdot\wt{\mb{M}}_V\xr\bigg)\nonumber\\
	&=\intx\bigg(\mu_0\bigg\{\left[\wt{\mb{M}}_V\cdot\bN_{\dot{\mb{r}}}\right]\mb{M}+\wt{\mb{M}}_V\times\left[\bN_{\dot{\mb{r}}}\times\mb{M}\right]\bigg\}\nonumber\\
	&\hspace{3.7cm}-\bigg\{\left[\mb{B}\cdot\bN_{\dot{\mb{r}}}\right]\wt{\mb{M}}_V+\mb{B}\times\left[\bN_{\dot{\mb{r}}}\times\wt{\mb{M}}_V\right]\bigg\}\bigg)\\
	&=\intx\left\{\mu_0\left[\bm{\theta}\xr\times\wt{\mb{M}}_V\xr\right]-\bN_{\dot{\mb{r}}}\left[\mb{B}\x\cdot\wt{\mb{M}}_V\xr\right]\right\}.
\end{align}
Combining these, we find that
\begin{align}
	\wt{\mb{p}}&=m\dot{\mb{r}}-\intx\bm{\theta}\rr\times\mb{B}\x-\bN_{\dot{\mb{r}}}\intx\mb{B}\x\cdot\wt{\mb{M}}_V\xr\\
	&\equiv m\dot{\mb{r}}-\bm{\Phi}_0\rr-\wt{\bm{\Phi}}\rr
\end{align}
as in Eqn.~(21a) in the main text.
\subsection*{B. Hamiltonian}
As usual we write the Hamiltonian in terms of the coordinates and momenta as in Eqn.~\eqref{eq:Hgen}
\begin{equation}
\wt{H}=\wt{\mb{p}}\cdot\dot{\mb{r}}-\frac{1}{2}m\dot{\mb{r}}^2+\intx\left(\mb{B}\x\cdot\dot{\wt{\mb{C}}}\x-\frac{1}{2\mu_0}\mb{B}^2\x+\frac{\varepsilon_0}{2}\mb{E}^2\x\right).\label{eq:HV1}
\end{equation}
We then define $\wt{\mb{X}}=\bm{\Phi}_0+\wt{\bm{\Phi}}$ and note the following three relations:
\begin{itemize}
	\item Using the definition $\dot{\wt{\mb{C}}}+\bN \wt{C}_0=\mu_0^{-1}\mb{B}-\wt{\mb{M}}$ and that $\dive\mb{B}=0$ leads to
	\begin{equation}
	\intx\mb{B}\x\cdot\wt{\mb{C}}\x=\intx\left(\frac{1}{\mu_0}\mb{B}^2\x-\mb{B}\x\cdot\wt{\mb{M}}\xr\right).
	\end{equation}
	\item The terms outside the volume integral can be written as
	\begin{equation}
	\wt{\mb{p}}\cdot\dot{\mb{r}}-\frac{1}{2}m\dot{\mb{r}}^2=\frac{1}{2m}\left(\wt{\mb{p}}^2-\wt{\mb{X}}^2\rr\right).
	\end{equation}
	\item We can rewrite
	\begin{equation}
	-\intx\mb{B}\x\cdot\mb{M}\xr=\frac{1}{m}\left(\wt{\mb{p}}\cdot\wt{\mb{X}}\rr+\wt{\mb{X}}^2\rr\right)-\dot{\mb{r}}\cdot\wt{\bm{\Phi}}\rr.
	\end{equation}
\end{itemize}
Inserting these three relations into Eqn.~\eqref{eq:HV1} leads to
\begin{equation}\label{eq:HnewVfix}
\wt{H}=\frac{1}{2m}\left(\wt{\mb{p}}+\wt{\mb{X}}\rr\right)^2+\frac{1}{2}\intx\left[\frac{1}{\mu_0}\mb{B}^2\x+\varepsilon_0\mb{E}^2\x\right]+\left(1-\dot{\mb{r}}\cdot\bN_{\dot{\mb{r}}}\right)\intx\mb{B}\x\cdot\wt{\mb{M}}_V\xr.
\end{equation}
The third term in Eqn.~\eqref{eq:HnewVfix} is a result of not being able to invert the matter canonical momentum for an expression for $\dot{\mb{r}}$ in terms of $\wt{\mb{p}}$. We therefore impose a constraint on the choice of $\mb{V}$ and $V_0$ such that:
\begin{equation}
\left(1-\dot{\mb{r}}\cdot\bN_{\dot{\mb{r}}}\right)\intx\mb{B}\x\cdot\wt{\mb{M}}_V\xr=0.
\end{equation}
This extra constraint on the transformation follows from
\begin{equation}
\wt{\bm{\Phi}}\rr=\Dfrac{}{\dot{\mb{r}}}\intx\mb{B}\x \cdot\wt{\mb{M}}_V\xr,
\end{equation}
being in general a complicated function of $\dot{\mb{r}}$. Therefore, it is not possible obtain for $\dot{\mb{r}}$ in terms of $\wt{\mb{p}}$ by inverting the canonical momentum Eqn.~(21a). As an example of an allowed transformation, consider $\mb{V}=\frac{1}{2}\mb{r}\times\mb{x}$ and $V_0=0$. For this transformation, we find that $\wt{\bm{\Phi}}=\frac{1}{2}\intx\mb{x}\times\mb{B}$ which means that we can write $\dot{\mb{r}}$ as a function of $\wt{\mb{p}}$ by inverting Eqn.~(21a). Importantly, this also means that the last term of Eqn.~\eqref{eq:HnewVfix} vanishes, and we arrive at Eqn.~(23) in the main text.

\section*{Supplementary Material 5: Table of experimental data points}\label{app:table}
\begin{table*}[ht]
\caption{Table showing the ranges of values used in Figures 2 with corresponding references. Much of the data rely on estimated parameters and so should be used as guide values. The data range with the asterisk has a much larger maximum value than expected, so should be taken with caution. }
	\begin{ruledtabular}
		\begin{tabular}{cccccccc}
			Experiment & $\eta$ & $\nicefrac{f}{\mathrm{eV}}$ & Dipole size / $\mu$m & $\nicefrac{g}{\nu}$ & $\nicefrac{g}{\omega_{10}}$ & References\\
			\colrule 
			\mc{Rb gas in}{optical cavity} & $2.0\times10^{-4}$ & \mc{$0.0045$}{$-0.19$} & $1.6\times10^{-4}$ & \mc{$2.48\times10^{-10}$}{$-1.42\times10^{-6}$} & \mc{$2.48\times10^{-10}$}{$-1.42\times10^{-6}$} & \cite{suleymanzade2019tunable,colombe2007strong,thompson2013coupling,tiecke2014nanophotonic} \\
			\mc{Quantum}{dot arrays} & \mc{$0.012$}{$-0.052$} &\mc{$0.023$}{$-0.034$}& \mc{$0.01$}{$-0.05$} &\mc{$5.3\times10^{-4}$}{$-0.0020$} & \mc{$6\times10^{-4}$}{$-0.0019$} & \cite{gerard1998enhanced,gerard1996quantum,gerard1998inas,gerard1998inas,gayral1999high,moreau2001single} \\
			\mc{Superconducting}{circuit} & \mc{$5.6\times10^{-4}$}{$-5.9\times10^{-4}$} &\mc{$8\times10^{-4}$}{$-0.029$}& \mc{$21.4$}{$-60$}& \mc{$0.073$}{$-1.3$} & \mc{$0.040$}{$-15.3^*$} & \cite{johansson2006vacuum,niemczyk2010circuit,forn2010observation,baust2016ultrastrong,yoshihara2017superconducting} \\
			\mc{Rare earth spins in}{microwave resonator} & $7.2\times10^{-4}$ & $3.3\times10^{-4}$ & $299.4$ & $0.15$ & $0.051$ & \cite{everts2019ultrastrong} \\
			\mc{Exciton}{polaritons} & \mc{$0.0091$}{$-0.0098$} & \mc{$0.11$}{$-0.51$} & \mc{$0.0070$}{$-0.0076$} & \mc{$0.0012$}{$-0.0060$} & \mc{$0.0012$}{$-0.0060$} & \cite{weisbuch1992observation,bloch1998giant} \\
			\mc{Exciton polaritons}{in dyes} & \mc{$6.0\times10^{-4}$}{$-0.0011$} & \mc{$44.94$}{$-484.5$} & \mc{$5\times10^{-4}$}{$-5.1\times10^{-4}$} & \mc{$0.035$}{$-0.037$} &\mc{$0.043$}{$-0.30$} & \cite{bellessa2004strong,wei2013adjustable,gambino2014exploring,kena2013ultrastrongly} \\
			\mc{Intersubband polaritons}{in quantum wells} & \mc{$3.0\times10^{-4}$}{$-0.0037$} & \mc{$4.7$}{$-98.6$} & \mc{$0.0075$}{$-0.032$} &\mc{$0.023$}{$-2.8$} & \mc{$0.032$}{$-4.0$} & \cite{dupont2003vacuum,dupont2007giant,todorov2010ultrastrong,delteil2012charge,askenazi2014ultra} \\
			\mc{Electron cyclotron}{resonances} & \mc{$1.7\times10^{-6}$}{$-2.0\times10^{-4}$} & \mc{$0.76$}{$-25.7$} & \mc{$0.073$}{$-0.19$} & \mc{$0.17$}{$-4.8$} & \mc{$0.058$}{$-1.4$} & \cite{muravev2011observation,scalari2012ultrastrong,maissen2014ultrastrong,bayer2017terahertz}
		\end{tabular}
	\end{ruledtabular}
\end{table*}

\noindent We have used various estimates when the dipole size has not been reported. In particular: 
\begin{itemize}
\item The dipole size $x_{10}$ is approximately a couple of Bohr radii $a_0$ for atomic species, whereas the effective Bohr radius $a_\text{eff} \simeq 1 \text{ }\mu\text{m}$ is used for Rydberg atoms. 
\item For quantum dots and quantum wells, the physical extent of the device is a good estimate for the dipole size. 
\item The estimates become slightly more complicated for superconducting circuits, but an approximate value can be extracted in two different ways, depending on the values reported: First from the definition of the dipole moment
\begin{align}
    x_{10} &\simeq |\mb{d}|/e \text{ with } |\mb{d}| = \hbar g/E_\text{vac}, \\
    &\Rightarrow x_{10}\simeq \hbar g/(e E_\text{vac}),
\end{align}
where $g$ is the interaction strength, and $E_\text{vac}$ is the electric field amplitude of the vacuum fluctuations. Electric field fluctuations can in turn be estimated by $E_\text{vac} \simeq (V_\text{vac}/\text{length of relevant region})$, with $V_\text{vac}$ being the fluctuations in the electric potential. Alternatively, it is possibly to define an analogous Bohr radius $a_\text{eff}$ through the resonance frequency, as in 
\begin{align}
    \omega_{10} &= 2\pi c/\lambda_{10} = 2\pi c \left(\frac{\alpha}{4\pi a_\text{eff}}\right),\\
    &\Rightarrow a_\text{eff} \simeq \frac{\alpha c}{2\omega_{10}},
\end{align}
where $\alpha$ is the fine-structure constant and where we have interpreted the transition wavelength $\lambda_{10} = 4\pi a_\text{eff}/\alpha$ as an effective Bohr wavelength. 
\item In dye-filled polariton cavities, the spatial extent of the dye molecule acts as a bound of the dipole size. 
\item In the case of electron cyclotron resonances, we can estimate the dipole size as 
\begin{align}
x_{10} \simeq l_0 \sqrt{\nu},    
\end{align}
where $l_0 = \sqrt{\hbar/(e B)}$ is the magnetic length for a given magnetic field strength $B$, and $\nu = \rho_{2\text{deg}}2\pi l_0^2$ is the filling factor in the system with $\rho_{2\text{deg}}$ being the electron density. Note that in some cases, the filling factor is directly reported.

\end{itemize}

\section*{Supplementary Material 6: Accuracy of few level truncation without resonance}\label{app:offres}
\begin{figure*}[ht]\centering
\includegraphics[width=0.7\textwidth]{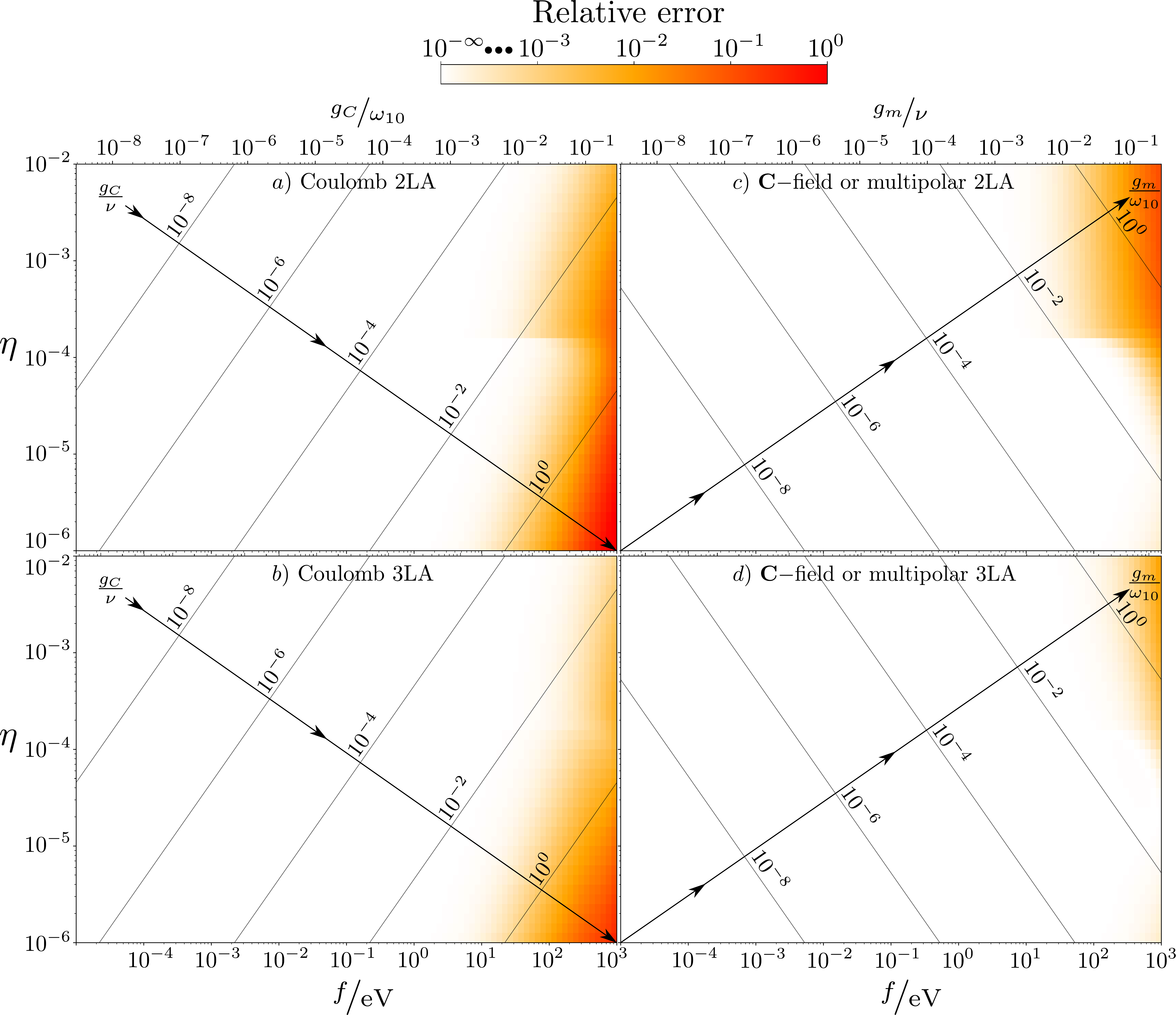}
\caption{As in Figure~2 in the main text, we plot the relative error in calculating the energy spacing between the two lowest eigenstates of the full Coulomb gauge and $\mb{C}$-field Hamiltonians within the two and three dipole-level approximations, using an infinite square well potential as $\widehat{U}_{\mathrm{ext}}$. The difference is that in this plot the lowest lying dipole transition is kept fixed at $1$~eV, and so, is not on resonance with the mode frequency which is varied along the $y$-axes. For all other details refer to Figure~2. } \label{fig:nonres}
\end{figure*}

\noindent Figure \ref{fig:nonres} is the same calculation as for Figure~2 in the main text, however, the splitting between the two lowest lying dipole levels is no longer kept resonant with the single photon mode $\nu$. Instead, we fix the length of the infinite square well such that the lowest lying dipole transition, $\omega_{10}=\epsilon_1-\epsilon_0=1~\mathrm{eV}$ is held constant. Outwith resonance, the light-matter coupling ratios with $\omega_{10}$ and $\nu$ in either gauge is not the same, and so the contours along the diagonal-axis and $x$-axis are different. Just as in Figure~2, even without resonance the $\mb{C}$-field calculations yield more accurate results for a given coupling strength than the Coulomb gauge, and also converge onto the exact answer quicker as the number of dipole levels in the calculation is increased. Note also that it is for large dipoles in the strong-coupling limit that the $\mb{C}$-field calculations become inaccurate, i.e. in the scenario when we expect the description of the polarisation field in terms of dipole degrees of freedom to become non-trivial.

\end{document}